\documentclass[]{pasj02}

\jyear{2025}
\Received{}
\Accepted{}

\usepackage{threeparttablex}
\usepackage{xspace}
\usepackage{url}
\usepackage{rotating}
\usepackage{booktabs} 
\usepackage{longtable}
\usepackage{natbib}

\usepackage{fix-cm}

\DeclareRobustCommand{\ion}[2]{\textup{#1\,\textsc{\lowercase{#2}}}}

\begin{document} 

\title{ 
Indication of the Less-ionized Clumpy Ultra-Fast Outflows in Seyfert Galaxies } 

\author{Takuya \textsc{Midooka}\altaffilmark{1,2,3}} %
\altaffiltext{1}{Institute of Space and Astronautical Science, Japan Aerospace Exploration Agency,
3-1-1 Yoshinodai, Chuo-ku, Sagamihara, Kanagawa 252-5210, Japan}
\altaffiltext{2}{Department of Astronomy, Graduate School of Science, The University of Tokyo, 7-3-1 Hongo, Bunkyo-ku, Tokyo 113-8654 Japan}
\altaffiltext{3}{Mercari, 6-10-1 Roppongi, Minato-ku, Tokyo 106-6118, Japan}

\author{Misaki \textsc{Mizumoto},\altaffilmark{4,}\footnotemark[*]}
\altaffiltext{4}{Science education research unit, University of Teacher Education Fukuoka, 
1-1 Akama-bunkyo-machi, Munakata, Fukuoka 811-4192, Japan}
\email{mizumoto-m@fukuoka-edu.ac.jp}

\author{Ken \textsc{Ebisawa}\altaffilmark{1}}

\KeyWords{galaxies: active --- galaxies: Seyfert --- X-rays: galaxies
}

\maketitle

\begin{abstract}
We present a systematic investigation of X‐ray spectral variability of Seyfert 1 galaxies using a ``spectral‐ratio model fitting'' technique, which we developed to estimate contribution of the putative clumpy absorbers to the spectral variations. Archival XMM–Newton observations of 12 active galactic nuclei were analyzed to constrain properties of these absorbers. Our analysis demonstrates that the soft X–ray variability is primarily governed by fluctuation of the partial covering fraction of mildly ionized clumpy clouds. In particular, for Mrk 335, PDS 456, and 1H 0707-–495, outflow velocities of the clumpy absorbers are constrained from the blue-shifts of the Fe-L edge structure. The blue-shifted Fe-L edge successfully reproduces the well-known complex spectral feature near 1 keV in 1H 0707--495, which was often explained by invoking an ad hoc absorption structure. Notably, the inferred outflow velocities of the clumpy absorbers are comparable to, or even exceed, those associated with the Ultra–Fast Outflows (UFOs). Furthermore, we found a positive correlation between the outflow velocities and the intrinsic X–ray fluxes in two of four data sets, and the remaining two datasets also agree with this positive correlation, which supports a radiative-driven wind scenario that the X-ray/UV emission from the central black holes is causing the UFOs and the outflowing clumpy absorbers. In addition, the line-driven acceleration is likely playing a significant role, since the line opacities of the clumpy absorbers are highly sensitive to the flux changes. These findings provide a robust observational support for the ``hot inner and clumpy outer wind'' paradigm, suggesting a common origin for both the UFOs and the clumpy absorbers. 
\end{abstract}


\section{Introduction}\label{sec:intro} 

Observational evidence has firmly established existence of the various types of the X-ray absorbers in Active Galactic Nuclei (AGNs), such as ionized Ultra-Fast Outflows (UFOs; e.g., \citealt{king03,pounds03,Reeves09}) and clumpy absorbers (e.g., \citealt{Tanaka04,miller08,turner09,sanfrutos2013,markowitz2024}). These absorbers are known to exhibit substantial temporal variability. Among them, clumpy absorbers are particularly notable for inducing significant variability in the soft X-ray band through changes of their covering fraction (e.g., \citealt{ lamer03, Behar03,gallo04,turner07,Iso16}). 

 Variability of the clumpy absorbers is consistent with theoretical models that predict instabilities in AGN winds. For example, the overlap of dense upstream regions with less dense downstream regions can trigger Rayleigh-Taylor instabilities (e.g., \citealt{mathews77}). Coupled with the radiation hydrodynamic instabilities, these mechanisms amplify clumpiness of the absorbers (e.g., \citealt{chelouche03,Takeuchi13,Takeuchi14}). From observational side, the ``hot inner and clumpy outer wind'' model \citep{Mizumoto19} provides a conceptual framework, suggesting that smooth UFOs dominate the inner region, while clumpy absorbers prevail in the outer region. This model successfully explains spectral variability of AGNs by accounting for distinct physical conditions in these regions and posits a physical connection between the UFOs and the clumpy absorbers.

Despite its conceptual utility, the standard X-ray spectral analysis often faces critical challenges. The parameters describing partial absorption by clumpy absorbers often degenerate with those of other spectral components, such as warm absorbers and the soft excess emission. This degeneracy complicates precise characterization and limits our ability to understand the dynamics and interactions of these absorbers within the AGN environment.

To address these limitations, \citet{Midooka22a} introduced a novel technique called ``spectral-ratio model fitting''. By normalizing intensity-sliced spectra to the brightest spectrum, this method isolates spectral variations caused by clumpy absorbers while eliminating nonvariable continuum and absorption features. 
\citet{midooka23} applied this technique to the narrow-line Seyfert 1 galaxy IRAS 13224--3809, and revealed that the soft X-ray spectral variations are primarily driven by changes in the covering fraction of the mildly ionized clumpy absorbers which exhibit high velocities comparable to those of the UFOs ($\sim0.2-0.3 \; c$). Furthermore, both velocities of the clumpy absorbers and the UFOs were found to increase with the intrinsic UV/X-ray luminosity, supporting their shared origin as the radiatively driven winds. This is an observational evidence that the clumps responsible for the partial absorption have the same origin as the radiation-driven UFOs, as expected by the hot inner and clumpy outer wind scenario \citep{Mizumoto19}.

Based on these studies, the aim of this paper is to investigate whether the correlation between the clumpy absorbers and the UFOs observed in IRAS 13224–3809 is found also in other AGNs. 
By applying the spectral-ratio model fitting technique to archival data from multiple AGNs, we examine whether the correlations between the clumpy absorbers, UFOs, and AGN luminosity are universal.

\section{Observations and data reduction}\label{Chapter3}

\subsection{Target selection}\label{sec:3.1}
The primary objective of this study is to constrain the outflow velocities of the clumpy absorbers in AGNs where UFO absorption lines have been detected. 
For this purpose, we selected AGN samples from the UFO catalogs compiled by \citet{Tombesi10a} and \citet{Igo20}, focusing on the UFOs detected with high significance (classified as ``Likely outflows'' in \citealt{Igo20}), where the total exposure time exceeds 200 ks. In this manner, in addition to IRAS 13224-3900, which was already analyzed in \citet{midooka23}, we selected eight AGNs for this study. To provide a comparative analysis, we also included four AGNs where no significant UFOs were detected in \citet{Igo20}, but presence of the clumpy absorbers was confirmed. The selected observation targets are summarized in Table~\ref{tab:obs_target}.

\begin{table*}
 \begin{threeparttable}
 \caption{Observation targets and their characteristics.} 
 \label{tab:obs_target}
\begin{tabular}{lrrrrrr}
\hline
1. Source & 2. AGN Class & 3. $z$ & 4. log $M_{\rm BH}$ ($M_{\odot}$) & ~~~~~~5. $v_\mathrm{UFO}$ (/c)&&\\\hline
\multicolumn{7}{c}{{\it UFO absorption detected in \citet{Igo20}}} \\
(IRAS~13224$-$3809) & NLS1 & 0.0658 & 6.0$\pm0.3$\tnote{R, a} & $0.238^{+0.003}_{-0.049}$\tnote{j} \\ 
1H~0707$-$495 & NLS1 & 0.04057 & 6.5$\pm0.2$\tnote{b} & $0.165^{+0.002}_{-0.051}$\tnote{j} \\
ESO~323$-$G77 & S1, NLS1 & 0.014904 & $\sim$7.12\tnote{D, c} & $0.085^{+0.006}_{-0.048}$\tnote{j} \\ 
MCG--6-30-15 & S1, NLS1 & 0.05708 & 6.30$^{+0.16}_{-0.24}$\tnote{R, d} & $0.08^{+0.02}_{-0.05}$\tnote{j} \\
Mrk~766 & NLS1 & 0.01271 & 6.82$^{+0.05}_{-0.06}$\tnote{R, d} & $0.08^{+0.02}_{-0.05}$\tnote{j} \\
NGC~4051 & NLS1 & 0.0023 & 5.89$^{+0.08}_{-0.15}$\tnote{R, d} & $0.061^{+0.022}_{-0.033}$\tnote{j} \\
NGC~7314 & NLS1 & 0.004771 & 5.9$\pm0.3$\tnote{V, e} & $0.04^{+0.02}_{-0.03}$\tnote{j} \\
PDS~456 & QSO & 0.184 & 9.3$\pm0.4$\tnote{E, f} & $0.255^{+0.049}_{-0.020}$\tnote{j} \\
Ton~S180 & NLS1 & 0.06168 & 7.3$\pm0.3$\tnote{E, g} & $0.35^{+0.05}_{-0.02}$\tnote{j} \\
\hline
\multicolumn{7}{c}{{\it UFO undetected in \citet{Igo20}, but clump features confirmed}} \\
Mrk~335 & NLS1 & 0.025 & 7.23$^{+0.04}_{-0.04}$\tnote{R, d}~~ & $0.051^{+0.053}_{-0.020}$\tnote{j'} \\
NGC~3783 & S1 & 0.009755 & 7.08$^{+0.05}_{-0.06}$\tnote{R, d}~~ & No detection \\ 
NGC~985 & S1 & 0.04271 & $\sim$8.2\tnote{h} & No detection \\
PG~1126$-$041 & S1 & 0.06 & $\sim$8.1\tnote{V, i} & 0.05\tnote{k} \\\hline 
 \end{tabular}
 \begin{tablenotes} 
  \item[1.] Source name.
  \item[2.] AGN class based on H$\beta$ line width and redshift. NLS1 means Narrow line Seyfert 1, S1 means Seyfert 1, and QSO means quasar.
  \item[3.] Cosmological redshift.
  \item[4.] The logarithm of the estimated black hole mass, mass estimate method (R: reverberation mapping, V: stellar velocity dispersion, E: empirical relation with 5100~\AA\ luminosity and BLR radius, D: Relation between dust size and BLR size), and references (a: \cite{Parker17}, b: \cite{Done16}, c: \cite{Gravity22}, d: \cite{Bentz15}, e: \cite{McHardy13}, f: \cite{Emmanoulopoulos14}, g: \cite{Turner02}, h: \cite{Vasudevan09}, i: \cite{Dasyra07})
  \item[5.] The detected UFO velocity and references (j: \cite{Igo20}, j': \cite{Igo20}, but low significance, k: \cite{Giustini11}).
 \end{tablenotes}    
 \end{threeparttable}
\end{table*}

\subsection{Data reduction}\label{sec:3.2}
XMM-Newton data are suitable for this study thanks to their high sensitivity, simultaneous optical and UV coverage, and long uninterrupted observation windows. Details of the XMM-Newton observations for the selected targets are provided in Table~\ref{tab:obs_log} and E-table 1.
As described in \citet{Midooka22a,midooka23}, the spectral-ratio model fitting technique is most effective for analyzing relatively short timescale (e.g., weeks) spectral variations, where longer-term variations can be neglected. Thus, observations taken within a few weeks were grouped and treated as single observation sets for spectral-ratio analysis. Observations separated by years were analyzed as independent data sets (e.g., Group 1, Group 2). Within each observation group, spectral-ratio analysis was performed.

For all the targets, the EPIC-pn detector data were utilized. Data reduction was carried out using the Science Analysis Software (SAS; \citealt{Gabriel2004}).
Good time intervals were identified by excluding the periods of the high background flaring, defined by the {\tt PATTERN==0} count rates greater than 0.4~counts~s$^{-1}$ in the 10--12 keV energy range.
The source and background events were extracted using circular regions with radii of 250 physical units (12.5$^{\prime\prime}$) and 1600 physical units (75$^{\prime\prime}$), respectively, from the same detector chip. The background region was carefully chosen to avoid the chip edges and the areas with high Cu background contamination. In addition to the pn data, the OM data were processed using the standard SAS routine {\tt omichain}, and the spectrum was extracted using {\tt om2pha}.


\begin{table*}
 \begin{threeparttable}
 \caption{\textit{XMM-Newton} (pn) observation logs (The full table is available only on the online edition as the supplementary data, E-table 1)} 
 \label{tab:obs_log}
\begin{tabular}{lrrrrr}
\hline
Source & Obs. Group & Obs. ID & Start Date & Exp. (ks)$^\mathrm{a}$ & count rate (cts s$^{-1}$)$^\mathrm{b}$ \\ 
\hline
IRAS~13224$-$3809 & 1 & 0780560101 & 2016-07-08 & 32.6 & 2.8 \\
 & & 0780561301 & 2016-07-10 & 127.4 & \\
 & & 0780561401 & 2016-07-12 & 81.8 & \\
 & & 0780561501 & 2016-07-20 & 130.4 & \\
 & & 0780561601 & 2016-07-22 & 130.4 & \\
 \hline
 \end{tabular}
 \begin{tablenotes} 
\item[a] Sum of all the good time intervals. For observations with unscheduled exposure, they are combined with scheduled ones.
\item[b] The mean one in the 0.3--10~keV \end{tablenotes}    
 \end{threeparttable}
\end{table*}


\section{Data Analysis and Results}\label{sec:3.3}
\subsection{Spectral ratios and their modeling}\label{sec:3.3.1}

\begin{figure}
 \centering
\includegraphics[width=0.95\columnwidth]{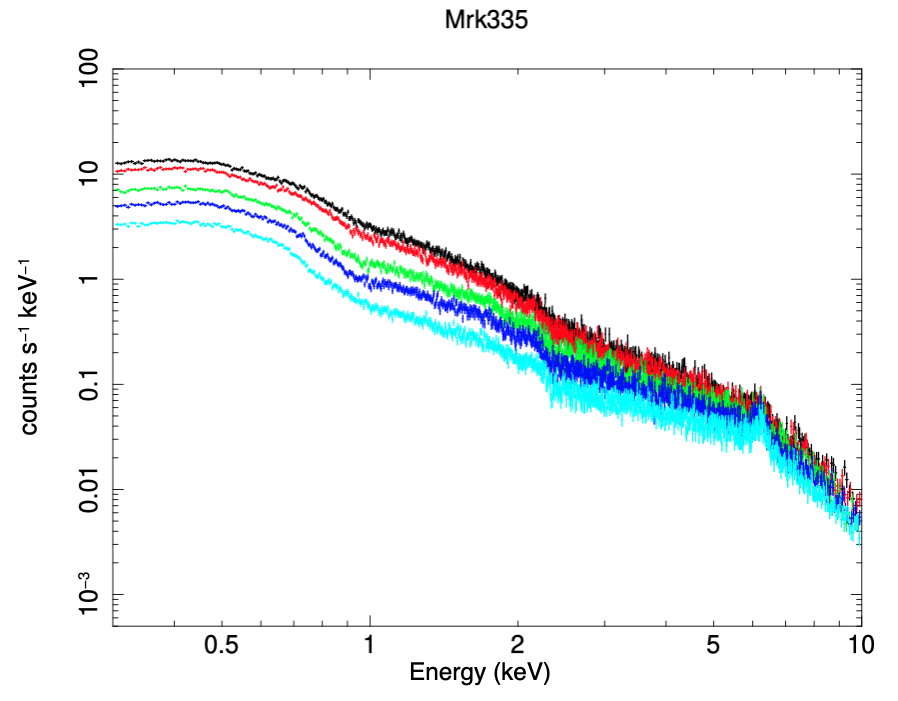}
\vspace{3mm}
\caption{  
An example of the intensity-sliced spectra of Mrk 335 at five different flux levels. The horizontal axis shows the X-ray energy (keV). All the figures are available only on the online edition as the
supplementary data (E-figure 1). \\
{Alt text: X-ray intensity-sliced spectra of the Seyfert galaxy Mrk 335 at five different flux levels. The horizontal axis represents the X-ray energy, ranging from 0.3 to 10 keV, and the vertical axis shows the photon flux in counts per second per keV on a logarithmic scale. Multiple spectra, color-coded, correspond to different intensity levels, with the brightest at the top and the dimmest at the bottom. The spectra reveal variations in X-ray flux at different energy levels, highlighting spectral changes due to clumpy absorbers.}}
\label{fig:others_slicedspec1}
  \end{figure}

To investigate the spectral variability of the selected AGNs, we created intensity-sliced spectra. All the 0.3--10.0 keV events were binned in 1~ks intervals, and the data were divided into five intensity levels (A--E), with A being the brightest and E the dimmest. This division ensured that the total number of photons in each intensity group was approximately equal. The resulting intensity-sliced spectra are shown in Figure \ref{fig:others_slicedspec1} and E-figure 1.

We then calculated the spectral ratios by dividing each intensity-sliced spectrum (B--E) by the brightest spectrum (A), following the methodology of \citet{Midooka22a}. This approach effectively isolates variations caused by changes in the covering fraction of the clumpy absorbers, while minimizing the impact of non-variable spectral components.

We generated spectral-ratio models in the 0.3--5.0 keV band using XSPEC (version 12.12.0; \citealt{Arnaud96}). We employed the zxipcf model \citep{Reeves08} to represent the photo-ionized plasma. This model includes three key parameters: the ionization parameter ($\xi$), the hydrogen column density ($N_\mathrm{H}$), and the partial covering fraction (CF). Additionally, we introduced the outflow velocity of the clumpy absorbers ($v_\mathrm{out}$) to account for the expected blue shift of the absorption features. Normalization of the continuum intensity was also included as a free parameter.
A table model for the spectral ratios was created by calculating the ratio of the absorbed model spectra to the unabsorbed one, maintaining the same continuum parameters. 
The brightest spectrum (A) is assumed to be unabsorbed.
This process was iterated across the entire parameter grid. The parameter range is listed in Table~\ref{tab:grid}.

\begin{table}
	\centering
	\caption{Parameter ranges for the spectral-ratio model.}
	\label{tab:grid}
\begin{tabular}{lccc} \hline 
Component & Minimum & Maximum & Number of steps \\ \hline 
CF & 0.0 & 1.0 & 21 \\ 
velocity [$c$] & $-$0.30 & 0.0 & 13 \\ 
$\log\xi$ & 0.0 & 3.0 & 13 \\ 
log $N_\mathrm{H}$ & 22.0 & 25.0 & 13 \\ 
\hline\end{tabular}
\end{table}

It should be noted that the model spectral ratios were generated by convolving the physical models with the detector response matrices (RMF and ARF), and then calculating the ratio of the absorbed model spectra to the unabsorbed one. We assumed that the shape of the underlying continuum (e.g., the power-law photon index) remains constant during the observation period of each group, allowing us to isolate the spectral changes induced by the variable absorber.

We tried to apply this method to  all the AGNs in Table \ref{tab:obs_target} except IRAS13224--3809; 
in total 12 sources are studied. 
For PG 1126$-$041 and ESO 323$-$G77, however, the spectral-ratio fitting could not be performed due to insufficient flux and a low signal-to-noise ratio. For Ton S180 and NGC 7314, no spectral ratio fitting was conducted either, as their intensity-sliced spectra hardly exhibited  time variation (E-figure 1). 

\begin{figure*}
  \centering
 \begin{minipage}{1.99\columnwidth}
   \centering
  \includegraphics[width=0.85\columnwidth]{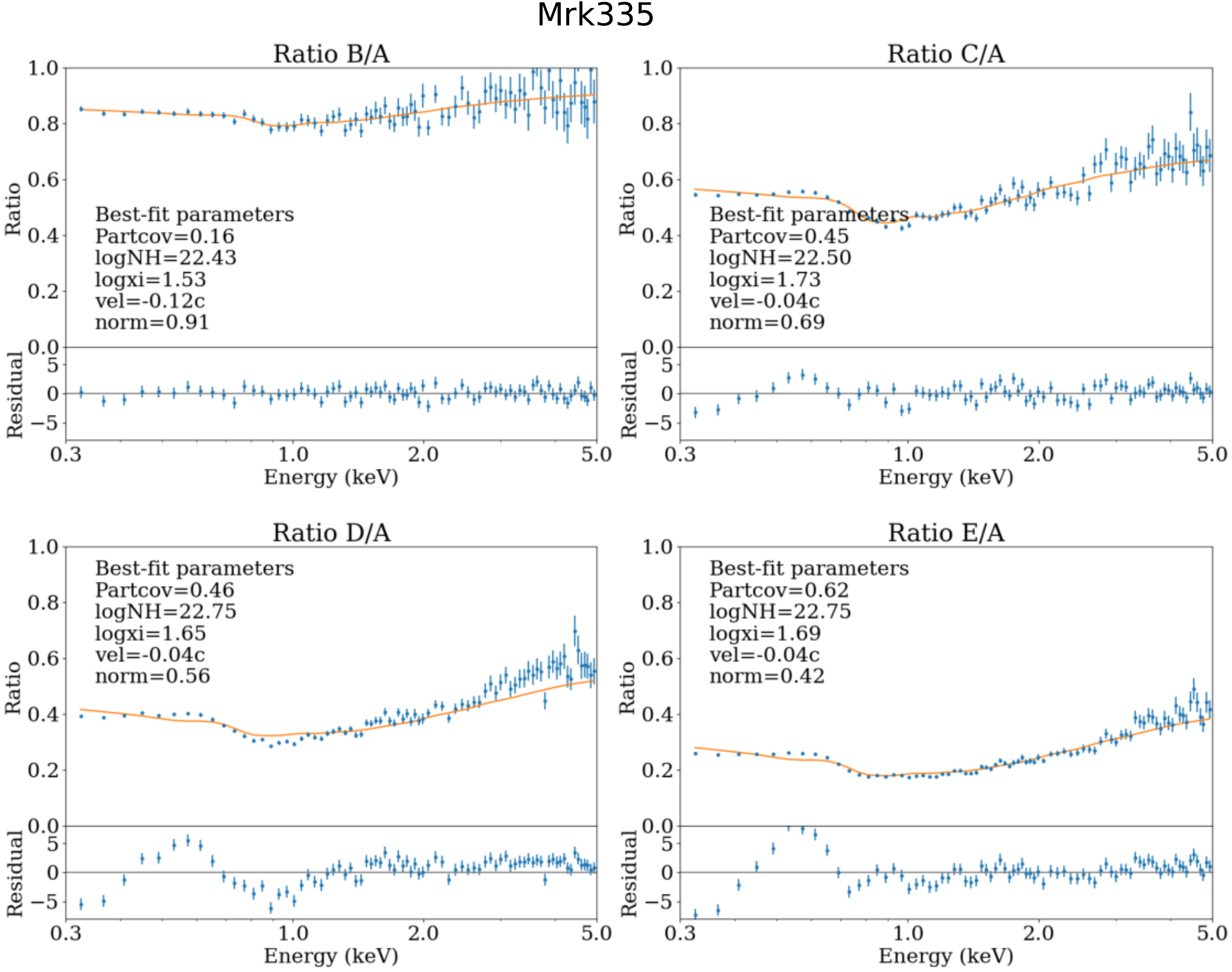}
  \end{minipage} 
\caption{  
Spectral ratio fitting results. Four clump parameters (CF, $\xi$, $N_\mathrm{H}$, and $v_\mathrm{out}$) and the normalization are free parameters. All the figures are available only on the online edition as the
supplementary data (E-figure 2).
\\
{Alt text: There are four sub figures and each sub figure has two panels. The upper panel shows the spectral ratio and the best fit model. The best-fit parameters are also listed. The lower panel shows the (data--model)/error values.
}}
\label{fig:eachfit}
\end{figure*}

The fitting results for the remaining eight objects --- Mrk 335, Mrk 766, NGC 985, NGC 4051, PDS 456, 1H 0707$-$495, MCG--6-30-15, and NGC 3783 --- are presented in Figure \ref{fig:eachfit} and E-figure 2.
We used the least chi-square method for the fitting.
We focus on the characteristic cliff or dip like features at $\sim$0.7--1 keV in the spectral ratios, 
which may be caused by the blue-shifted Fe-L edge.
For Mrk 335, PDS 456, and 1H 0707$-$495, the spectral ratios exhibit such structures around 1~keV, where the model ratios fit well with the expected blue-shifts. 
In contrast, the spectral ratios of Mrk 766, NGC 985, NGC 4051, MCG--6-30-15, and the 2001 observation of NGC 3783 do not show such characteristic features at $\sim$1 keV, showing no evidence of the clumpy outflows. 
We notice that the spectral ratio of NGC 3783 in 2016 shows a hump-like structure around 1~keV, but this is likely due to the constant scattered emission from distant gas \citep{Mao19, Marco20}.
 In the next section and on, we will focus on Mrk 335, PDS 456, and 1H 0707$-$495, which show evidence of the clumpy outflows.

\subsection{Simultaneous spectral-ratio model fitting}\label{sec:4.1}
We conducted a detailed analysis of the three objects (Mrk~335, PDS~456, and 1H~0707$-$495) that exhibited distinct dip structures in their spectral-ratios.
These dip features are considered strong evidence of the outflowing partial covering clouds. A precise analysis of these sources is expected to provide further constraints on the clump velocities and shed light on their physical origins.

In the previous subsection (section \ref{sec:3.3.1}), each spectral ratio was individually fitted. The fitting results revealed that $N_\mathrm{H}$ and $\xi$ of the clumpy absorbers did not show significant variations across different fluxes for each target (Figure \ref{fig:eachfit} and E-figure 2). Consequently, we
next perform simultaneous fitting for all the flux levels with these parameters tied. 

In the simultaneous fitting, we used a Markov Chain Monte Carlo (MCMC) approach using the ``emcee'' package \citep{Foreman-Mackey13} to determine the posterior distribution of the best-fit spectral parameters.
The initial parameter values were set on the basis of the best-fit results obtained from the chi-square fitting. A uniform prior distribution was applied within the predefined parameter ranges. After discarding the initial 5,000 steps to exclude the burn-in phase, an additional 10,000 steps were explored using 1,000 separate chains (walkers).


We successfully performed the MCMC parameter estimation for all the spectral ratios of the three targets. The spectral ratio fits and the corresponding best fit models are shown in Figure~\ref{fig:simulfit}. The best-fit parameters are indicated in Table \ref{tab:simul}.
Our findings indicate that, for all the three objects, the partial covering fraction decreases while the normalization increases as the X-ray flux increases. The outflow velocities of the clumpy absorbers in PDS 456 and 1H 0707$-$495 were in the range of 0.2--0.45~$c$, which are {\em comparable to the previously reported UFO velocities} (Table~\ref{tab:obs_target}). For Mrk~335, where the UFO velocity is lower ($\sim$0.05~$c$), the derived clump velocity is also {\em comparable to the UFO velocities}, ranging from 0.02 to 0.07~$c$. Furthermore, {\em a positive correlation was observed between the outflow velocities and the X-ray fluxes}, reinforcing the hypothesis that the clumpy absorbers and the UFOs share a common physical origin.
These trends are consistent with the results of IRAS~13224$-$3809 \citep{midooka23}.

To assess the reliability of the parameters, we examined the corner plots (Figure \ref{fig:MCMCcorner} and E-figure 3). Although a correlation between the partial covering fraction and the normalization is observed, as is often the case with partial covering models, the posterior distributions for both parameters are well-constrained. This confirms that the parameter coupling does not hinder accurate estimation of the covering fraction and velocity.

We also examined whether alternative models could explain the 0.7--1 keV feature. One possibility is the presence of an additional constant soft X-ray emission component, such as soft excess or scattered emission, which could dilute the variability and create an apparent dip in the spectral ratios. If the feature were caused by such a static component, the energy of the dip or edge would remain fixed regardless of the source flux. However, by calculating the spectral ratios, we found that the characteristic energy of the absorption feature systematically shifts to higher energies as the flux increases (i.e., the outflow velocity increases). This observed energy shift disfavors the constant component scenario. As demonstrated in the model-independent analysis of IRAS 13224–3809 \citep{midooka23}, this systematic blueshift can be a unique signature of the accelerating outflowing absorbers, supporting the variable wind scenario over the static alternative models.

\begin{figure*}
  \centering
  \includegraphics[width=1.6\columnwidth]{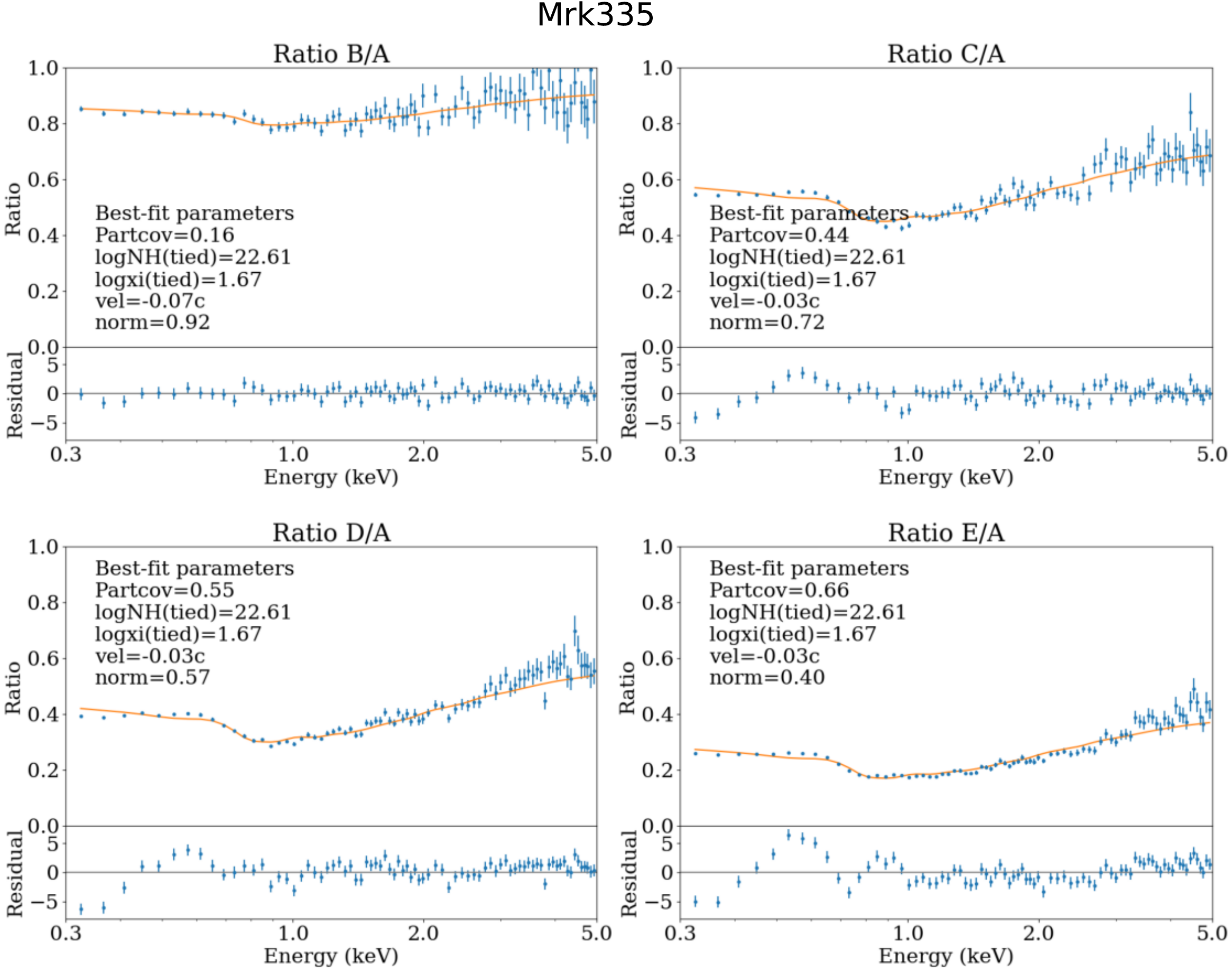}
  \includegraphics[width=1.6\columnwidth]{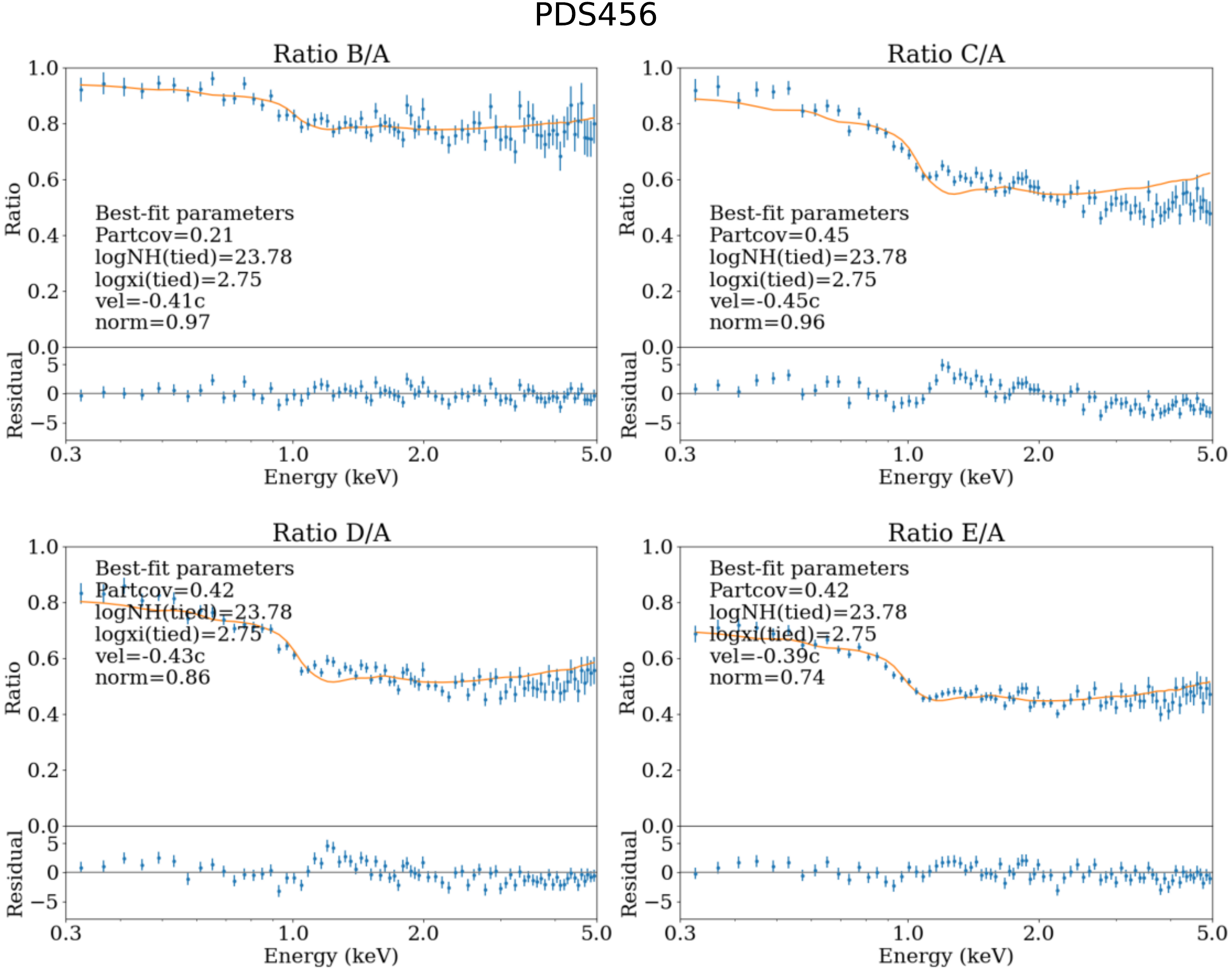}
  
  \caption{Simultaneous ratio fitting results.\\
{Alt text: Same as Figure \ref{fig:eachfit}, but for the simultaneous fitting.}} \label{fig:simulfit}
\end{figure*}

\addtocounter{figure}{-1}
\begin{figure*}
  \centering
  \includegraphics[width=1.6\columnwidth]{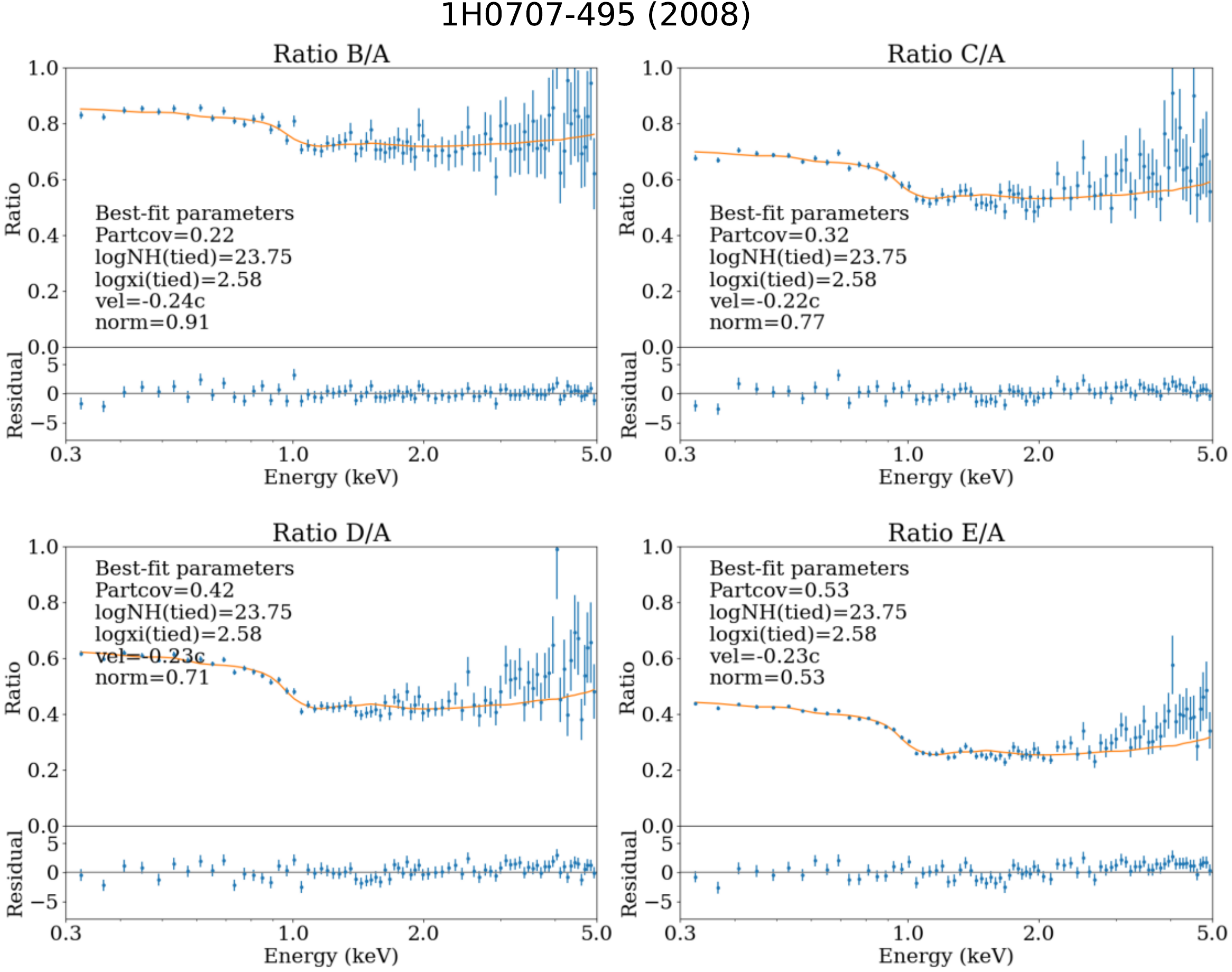}
  \includegraphics[width=1.6\columnwidth]{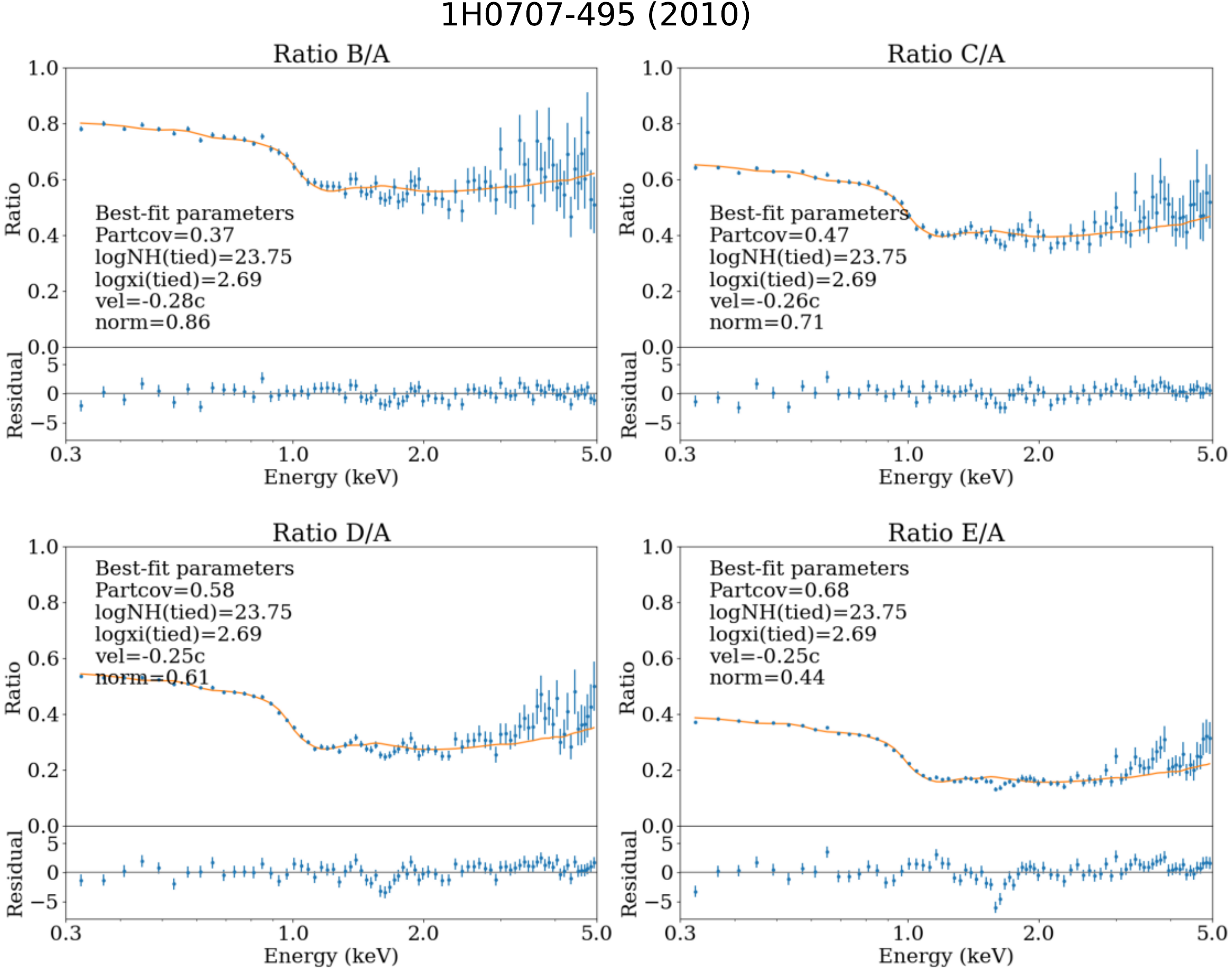}
  \caption{{\it Continued.}}
\end{figure*}

\begin{table*}
  \caption{Best-fit parameters determined by the simultaneous fitting, where $N_\mathrm{H}$ and $\xi$ are tied for all the spectral ratios.}
 \label{tab:simul}
\scalebox{1}{
  \begin{tabular}{cccccc}\hline
Mrk 335 &Parameters & B & C & D & E \\ \hline
& CF &$	0.16	\pm	0.02	$& $	0.436	\pm	0.010	$& $	0.549	\pm	0.008	$& $	0.665	\pm	0.007	$\\									
& $v_\mathrm{out}\,(/c)$&$	-0.10	\pm	0.04	$& $	-0.023	\pm	0.011	$& $	-0.020	\pm	0.008	$& $	-0.028	\pm	0.002	$\\					
& const &$	0.93	\pm	0.02	$& $	0.722	\pm	0.012	$& $	0.585	\pm	0.011	$& $	0.393	\pm	0.009	$\\				
& $N_\mathrm{H}\,(10^{23}\,\mathrm{cm}^{-2}$) & \multicolumn{4}{c}{$0.42\pm0.02$ (tied)}\\
& $\log\xi$ & \multicolumn{4}{c}{$1.67\pm0.02$ (tied)}\\
\hline
PDS 456 &Parameters & B & C & D & E \\ \hline
& CF &$	0.25	\pm	0.03	$& $	0.49	\pm	0.02	$& $	0.47	\pm	0.02	$& $	0.47	\pm	0.02	$\\	
& $v_\mathrm{out}\,(/c)$&$	-0.393	\pm	0.019	$& $	-0.463	\pm	0.012	$& $	-0.412	\pm	0.011	$& $	-0.377	\pm	0.008	$\\	
& const &	$	0.992	\pm	0.010	$& $	0.991	\pm	0.014	$& $	0.92	\pm	0.02	$& $	0.81	\pm	0.03	$\\	
& $N_\mathrm{H}\,(10^{23}\,\mathrm{cm}^{-2}$) & \multicolumn{4}{c}{$6.9\pm0.3$ (tied)}\\
& $\log\xi$ & \multicolumn{4}{c}{$2.75\pm0.02$ (tied)}\\
\hline
1H 0707--495 (2008) &Parameters & B & C & D & E \\ \hline
& CF &$	0.23	\pm	0.03	$& $	0.33	\pm	0.03	$& $	0.43	\pm	0.03	$& $	0.54	\pm	0.03	$\\			
& $v_\mathrm{out}\,(/c)$&$	-0.23	\pm	0.02	$& $	-0.222	\pm	0.014	$& $	-0.229	\pm	0.011	$& $	-0.232	\pm	0.007	$\\		
&const&	$	0.888	\pm	0.015	$& $	0.77	\pm	0.04	$& $	0.71	\pm	0.05	$& $	0.53	\pm	0.04	$\\		
& $N_\mathrm{H}\,(10^{23}\,\mathrm{cm}^{-2}$) & \multicolumn{4}{c}{$5.74\pm0.11$ (tied)}\\
& $\log\xi$ & \multicolumn{4}{c}{$2.7\pm0.3$ (tied)}\\
\hline
1H 0707--495 (2010) &Parameters & B & C & D & E \\ \hline
& CF &$	0.39	\pm	0.02	$& $	0.48	\pm	0.02	$& $	0.593	\pm	0.019	$& $	0.691	\pm	0.016	$\\		
& $v_\mathrm{out}\,(/c)$&	$	-0.280	\pm	0.009	$& $	-0.259	\pm	0.006	$& $	-0.252	\pm	0.005	$& $	-0.255	\pm	0.004	$\\
& const &	$	0.86	\pm	0.02	$& $	0.73	\pm	0.03	$& $	0.63	\pm	0.04	$& $	0.46	\pm	0.03	$\\	
& $N_\mathrm{H}\,(10^{23}\,\mathrm{cm}^{-2}$) & \multicolumn{4}{c}{$5.72\pm0.07$ (tied)}\\
& $\log\xi$ & \multicolumn{4}{c}{$2.7\pm0.3$ (tied)}\\
\hline
\end{tabular}
}
\end{table*}

\begin{figure*}
  \centering
  \includegraphics[width=1.6\columnwidth]{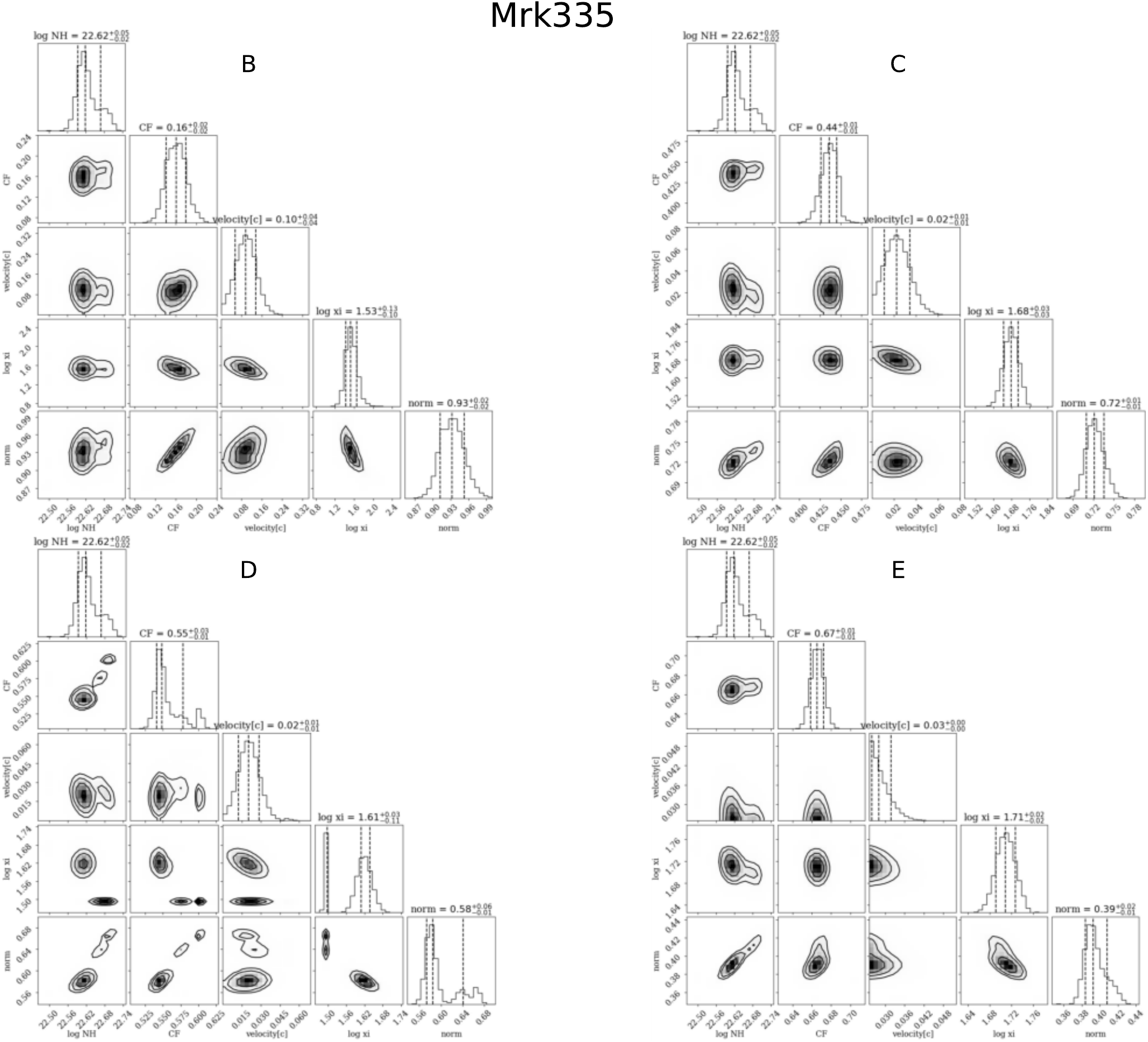}
  
  \caption{
Corner plots of the MCMC parameter estimation for all the spectral ratios. All the figures are available only on the online edition as the
supplementary data (E-figure 3).\\
{Alt text: The corner plot has five columns, log NH, CF, velocity, logxi, and norm. A contour plot for each parameter set and probability distribution for each parameter are shown.}}\label{fig:MCMCcorner}
\end{figure*}

\subsection{Spectral fitting in 0.3--10~keV}\label{sec:4.2}

To simultaneously constrain the outflow velocities of the clumpy absorbers ($v_\mathrm{out}$) and the UFO velocities ($v_\mathrm{UFO}$), spectral fitting was performed in the 0.3--10 keV energy band using a partial absorption model. Since the equivalent width of the UFO absorption in Mrk~335 was marginal to reliably estimate the velocity, spectral analysis was performed only for the remaining two objects, PDS~456 and 1H~0707$-$495, where UFO features were significantly detected.


The spectral data were rebinned to ensure at least 30 counts per energy bin. The spectral model used in XSPEC is expressed as follows:
\begin{equation} 
  {\tt phabs}*{\tt zxipcf}*({\tt powerlaw} + {\tt diskbb})*{\tt kabs},
  \end{equation}
which physically represents:
\begin{itemize}
  \item \textbf{ISM absorption (phabs):} Interstellar absorption was initially fixed to the Galactic value provided by \citet{Bekhti16}, with the cosmic abundances in \citet{Wilms00a}.
  \item \textbf{Partial covering absorption (zxipcf):} This component accounts for the clumpy absorbers with three key parameters: ionization parameter $\xi$, hydrogen column density $N_\mathrm{H}$, and partial covering fraction (CF).
  \item \textbf{Continuum emission:} Modeled with a power-law component representing the hot corona and a multicolor blackbody component for the optically thick accretion disk \citep{Mitsuda84, Makishima86}.
  \item \textbf{UFO absorption (kabs):} This model (\citealt{Ueda04}, updated by \citealt{Tomaru20}) describes the \ion{Fe}{XXV} and \ion{Fe}{XXVI} absorption lines, including the ion column density $N_\mathrm{ion}$, turbulent velocity $v_\mathrm{turb}$, and redshift $z$ due to the line-of-sight velocity.
\end{itemize}

Initial fitting parameters were set on the basis of the results of the spectral-ratio fitting. Both the continuum and the absorption parameters were allowed to vary during the fitting process. Since it is hard to simultaneously constrain the ionization parameter and the turbulent velocity from the absorption line profiles above 8 keV, $v_\mathrm{turb}$ was fixed at 10,000 km~s$^{-1}$, and the fitting was carried out under the assumption that the UFO absorption lines originate from either He-like or H-like Fe ions.
In the spectral fitting process, the K$\alpha$ and K$\beta$ absorption lines were included as a pair, the K$\beta$ line parameters determined independently from the K$\alpha$ lines. Up to three K$\alpha$ lines were required to fit the UFO features, indicating the presence of multiple velocity components.

\begin{figure*} 
\centerline{\includegraphics[width=1.95\columnwidth]{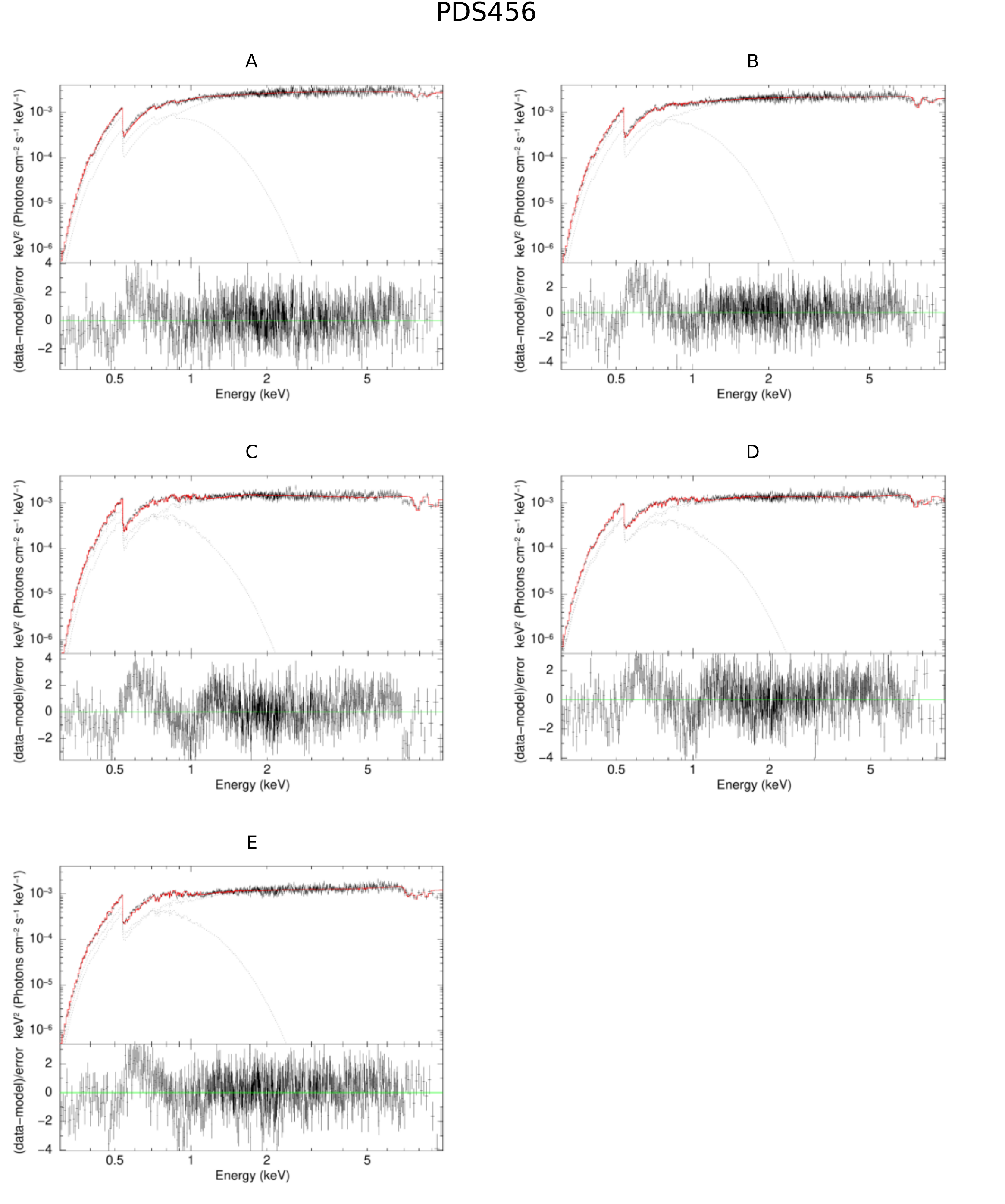}} 
	  \caption{Spectral fitting results of all the spectra of PDS 456 in 0.3--10~keV. The dotted lines in black show {\tt powerlaw} and {\tt diskbb}, respectively. The red line represents the best-fit model, including at maximum three {\tt kabs} lines around 7--10~keV. The lower panel shows the residuals of the model fitting.\\
{Alt text: There are 5 figures. The upper panel shows the $\nu F_\nu$ plot with the best fit model. The lowe panel shows (data--model)/error.}}
	  \label{fig:PDS456_specfit}
\end{figure*}

\begin{figure*} 
\centerline{\includegraphics[width=1.95\columnwidth]{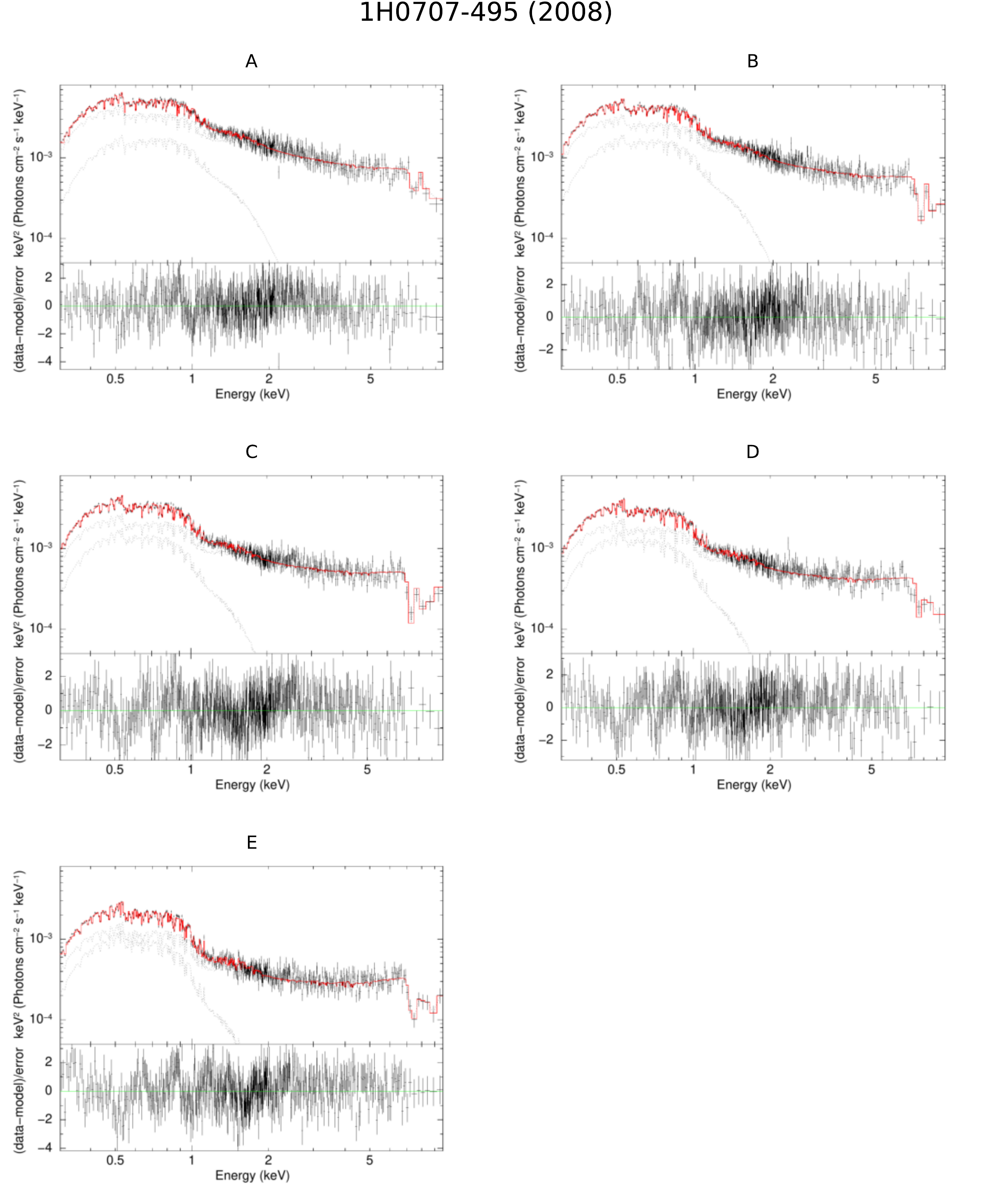}} 
	  \caption{The same as Figure~\ref{fig:PDS456_specfit}, but for 1H~0707$-$495 in 2008.\\
{Alt text: There are 5 figures.}}
	  \label{fig:1H0707_specfit_2008}
\end{figure*}

\begin{figure*} 
\centerline{\includegraphics[width=1.95\columnwidth]{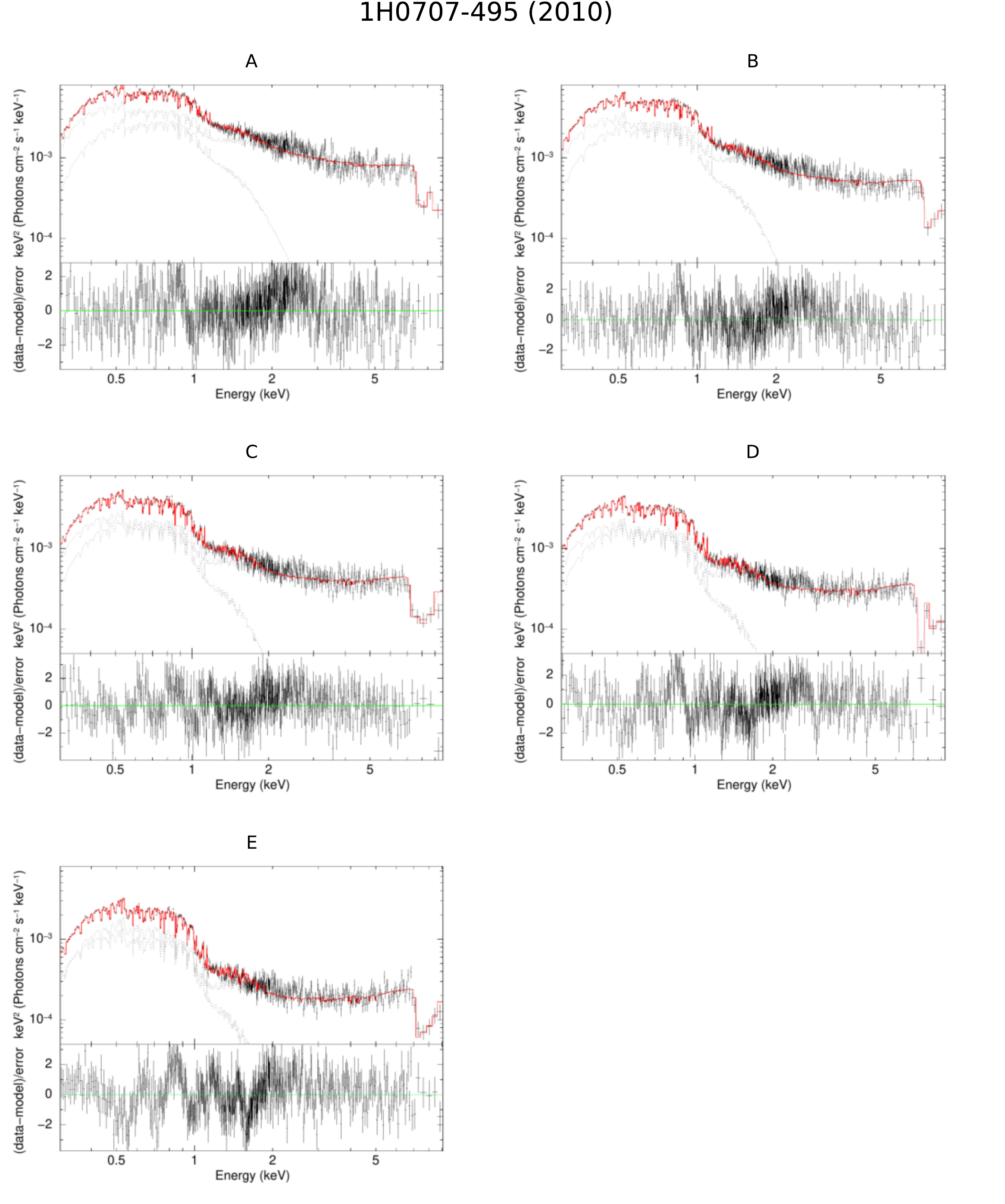}} 
	  \caption{The same as Figure~\ref{fig:PDS456_specfit}, but for 1H~0707$-$495 in 2010.\\
{Alt text: There are 5 figures.}}
	  \label{fig:1H0707_specfit_2010}
\end{figure*}

The best-fit results for PDS~456 and 1H~0707$-$495 are shown in Figures~\ref{fig:PDS456_specfit}--\ref{fig:1H0707_specfit_2010}, and the corresponding best-fit parameters are summarized in Tables~\ref{tab:best_specfit_PDS456}--\ref{tab:best_specfit_1H0707_2010}. 
The hydrogen column density ($N_\mathrm{H}$) of {\tt phabs} was allowed to vary, as it exhibited slight deviation from the value estimated by \citet{Bekhti16}. 
As indicated by the relatively large residuals in the spectral ratios of the dimmer states (e.g., see Figure~\ref{fig:simulfit}), the power-law photon index ($\Gamma$) of 1H~0707$-$495 was found to become steeper with increasing X-ray flux, which is consistent with the so-called ``softer-when-brighter'' scenario \citep{boller2021,parker21a}. 
For PDS~456, ionized partial absorption was not required for the spectrum A, i.e., CF$=0$. In contrast, for the spectrum A of 1H~0707$-$495, the partial covering component was clearly necessary (CF $>$ 0; see the middle and lower panels of Figure~\ref{fig:agnslim}), and the fit was performed with the clump parameters fixed to those obtained from the spectral-ratio fitting.
It should be reminded that the partial covering fractions obtained from the spectral-ratio fitting of 1H~0707$-$495
are underestimated, because the denominator (the spectrum A) was assumed completely uncovered  (see Figure~\ref{fig:simulfit}, Tables~\ref{tab:simul}, \ref{tab:best_specfit_1H0707_2008}, and \ref{tab:best_specfit_1H0707_2010}). 

\begin{table*}
  \caption{Best-fit parameters determined by the spectral fitting of PDS 456 in 0.3--10.0~keV. 
 The {\tt zxipcf} parameters are fixed to the values determined by the ratio fitting.
 Fitting was made assuming that the UFO absorption lines are produced by the He-like Fe.
 The parameters except {\tt kabs} are determined assuming the He-like Fe. Hereafter the error ranges correspond to the 90\% confidence level.}
 \label{tab:best_specfit_PDS456}
\scalebox{1}{
  \begin{tabular}{ccccccc}\hline
Components & Parameters & A & B & C & D & E \\\hline
{\tt phabs} & $N_\mathrm{H}$ (10$^{21}$cm$^{-2}$) & $3.7^{+0.2}_{-0.2}$ & $3.8^{+0.2}_{-0.2}$ & $4.1^{+0.2}_{-0.2}$ & $3.5^{+0.2}_{-0.2}$ & $3.6^{+0.2}_{-0.2}$ \\
{\tt zxipcf} & $N_\mathrm{H}$ (10$^{23}$cm$^{-2}$; fixed) & --- & 6.9 & 6.9 & 6.9 & 6.9 \\
 & log $\xi$ (fixed) & --- & 2.75 & 2.75 & 2.75 & 2.75 \\
 & CF (fixed) & 0 & 0.25 & 0.49 & 0.47 & 0.47 \\
 & $v_\mathrm{out}$ (/c; fixed) & --- & $-0.39$ & $-0.46$ & $-0.41$ & $-0.38$ \\ 
{\tt powerlaw} & $\Gamma$ & $2.06^{+0.02}_{-0.02}$ & $2.16^{+0.02}_{-0.02}$ & $2.42^{+0.02}_{-0.02}$ & $2.27^{+0.02}_{-0.02}$ & $2.20^{+0.02}_{-0.02}$ \\ 
 & norm (10$^{-3}$) & $3.2^{+0.1}_{-0.1}$ & $3.3^{+0.1}_{-0.1}$ & $4.0^{+0.1}_{-0.1}$ & $3.1^{+0.1}_{-0.1}$ & $2.6^{+0.1}_{-0.1}$ \\ 
{\tt diskbb} & $kT_{\rm in}$ (eV) & $0.16^{+0.01}_{-0.01}$ & $0.15^{+0.01}_{-0.01}$ & $0.13^{+0.00}_{-0.00}$ & $0.15^{+0.01}_{-0.01}$ & $0.15^{+0.01}_{-0.01}$ \\ 
 & norm (10$^{2}$) & $19^{+9}_{-6}$ & $26^{+12}_{-9}$ & $79^{+35}_{-25}$ & $16^{+8}_{-6}$ & $15^{+7}_{-5}$ \\ 
{\tt kabs} & $N_{\rm \ion{Fe}{XXV}}$ (10$^{18}$cm$^{-2}$) & $1.2^{+0.5}_{-0.5}$ & $2.5^{+0.7}_{-0.6}$ & $3.4^{+0.8}_{-0.7}$ & $3.1^{+0.8}_{-0.7}$ & $2.4^{+0.6}_{-0.6}$ \\ 
 & $z$ & $-0.67^{+0.01}_{-0.01}$ & $-0.31^{+0.01}_{-0.01}$ & $-0.33^{+0.01}_{-0.01}$ & $-0.29^{+0.01}_{-0.01}$ & $-0.26^{+0.01}_{-0.01}$ \\ 
{\tt kabs} & $N_{\rm \ion{Fe}{XXV}}$ (10$^{18}$cm$^{-2}$) & $0.9^{+0.6}_{-0.6}$ & $1.7^{+0.9}_{-0.7}$ & $2.2^{+1.1}_{-0.9}$ & $2.6^{+0.8}_{-0.7}$ & $3.1^{+0.7}_{-0.7}$ \\ 
 & $z$ & $-0.40^{+0.02}_{-0.02}$ & $-0.46^{+0.02}_{-0.01}$ & $-0.45^{+0.02}_{-0.02}$ & $-0.36^{+0.02}_{-0.02}$ & $-0.67^{+0.01}_{-0.01}$ \\ 
{\tt kabs} & $N_{\rm \ion{Fe}{XXV}}$ (10$^{18}$cm$^{-2}$) & --- & --- & --- & --- & $2.17^{+0.79}_{-0.70}$ \\ 
 & $z$ & --- & --- & --- & --- & $-0.40^{+0.01}_{-0.01}$ \\ 
 \hline
 $\chi^2$/dof & & 1071.2/979 & 1094.9/939 & 1190.5/870 & 1077.6/907 & 1122.8/929 \\\hline
\end{tabular}
}
\vspace{0.3cm}

  \caption{The same as Table~\ref{tab:best_specfit_PDS456} for 1H~0707$-$495 in 2008, but assuming only the He-like iron absorption line. CF parameter of the spectrum A is free to vary, unlike that of the other objects. 
  }
 \label{tab:best_specfit_1H0707_2008}
\scalebox{1}{
  \begin{tabular}{ccccccc}\hline
Components & Parameters & A & B & C & D & E \\\hline
{\tt phabs} & $N_\mathrm{H}$ (10$^{20}$cm$^{-2}$) & $8.0^{+0.7}_{-0.4}$ & $8.6^{+0.5}_{-0.5}$ & $8.0^{+0.5}_{-0.4}$ & $7.2^{+0.5}_{-0.5}$ & $6.5^{+0.5}_{-0.5}$ \\
{\tt zxipcf} & $N_\mathrm{H}$ (10$^{23}$cm$^{-2}$; fixed) & 6.9 & 6.9 & 6.9 & 6.9 & 6.9 \\
 & log $\xi$ (fixed) & 2.7 & 2.7 & 2.7 & 2.7 & 2.7 \\
 & CF & $0.52^{+0.02}_{-0.03}$ & $0.56^{+0.02}_{-0.02}$ & $0.60^{+0.02}_{-0.02}$ & $0.63^{+0.02}_{-0.02}$ & $0.70^{+0.02}_{-0.02}$ \\
 & $v_\mathrm{out}$ (/c) & $-0.29^{+0.01}_{-0.01}$ & $-0.28^{+0.01}_{-0.00}$ & $-0.27^{+0.00}_{-0.00}$ & $-0.27^{+0.00}_{-0.00}$ & $-0.26^{+0.00}_{-0.00}$ \\ 
{\tt powerlaw} & $\Gamma$ & $2.79^{+0.02}_{-0.02}$ & $2.77^{+0.02}_{-0.03}$ & $2.71^{+0.03}_{-0.03}$ & $2.72^{+0.03}_{-0.03}$ & $2.66^{+0.03}_{-0.04}$ \\ 
 & norm (10$^{-3}$) & $4.4^{+0.2}_{-0.2}$ & $3.4^{+0.2}_{-0.2}$ & $2.8^{+0.2}_{-0.2}$ & $2.4^{+0.2}_{-0.2}$ & $1.7^{+0.1}_{-0.1}$ \\ 
{\tt diskbb} & $kT_{\rm in}$ (eV) & $0.23^{+0.01}_{-0.01}$ & $0.21^{+0.01}_{-0.01}$ & $0.20^{+0.01}_{-0.01}$ & $0.19^{+0.01}_{-0.00}$ & $0.18^{+0.00}_{-0.00}$ \\ 
 & norm (10$^{2}$) & $1.6^{+0.6}_{-0.5}$ & $2.5^{+0.6}_{-0.6}$ & $2.9^{+0.7}_{-0.6}$ & $3.5^{+0.8}_{-0.7}$ & $2.9^{+0.7}_{-0.5}$ \\ 
{\tt kabs} & $N_{\rm \ion{Fe}{XXV}}$ (10$^{18}$cm$^{-2}$) & $4^{+2}_{-1}$ & $7^{+2}_{-1}$ & $8^{+3}_{-2}$ & $7^{+3}_{-2}$ & $7^{+2}_{-1}$ \\ 
 & $z$ & $-0.14^{+0.01}_{-0.01}$ & $-0.14^{+0.01}_{-0.01}$ & $-0.15^{+0.01}_{-0.01}$ & $-0.18^{+0.01}_{-0.01}$ & $-0.13^{+0.01}_{-0.01}$ \\ 
{\tt kabs} & $N_{\rm \ion{Fe}{XXV}}$ (10$^{18}$cm$^{-2}$) & $15^{+5}_{-4}$ & $11^{+6}_{-5}$ & $10^{+9}_{-4}$ & $9^{+9}_{-4}$ & $3^{+1}_{-1}$ \\ 
 & $z$ & $-0.27^{+0.01}_{-0.01}$ & $-0.25^{+0.01}_{-0.01}$ & $-0.25^{+0.01}_{-0.01}$ & $-0.29^{+0.01}_{-0.01}$ & $-0.20^{+0.02}_{-0.02}$ \\ 
{\tt kabs} & $N_{\rm \ion{Fe}{XXV}}$ (10$^{18}$cm$^{-2}$) & --- & --- & ---& ---& $4^{+3}_{-2}$ \\ 
 & $z$ & ---& ---& ---& ---& $-0.28^{+0.02}_{-0.02}$ \\ \hline
$\chi^2$/dof & & 587.1/515 & 596.1/525 & 568.9/525 & 573.7/507 & 688.6/517 \\\hline
\end{tabular}
}

\vspace{0.3cm}
\caption{The same as Table~\ref{tab:best_specfit_1H0707_2008} for 1H~0707$-$495 in 2010.}
 \label{tab:best_specfit_1H0707_2010}
\scalebox{1}{
  \begin{tabular}{ccccccc}\hline
Components & Parameters & A & B & C & D & E \\\hline
{\tt phabs} & $N_\mathrm{H}$ (10$^{20}$cm$^{-2}$) & $9.0^{+0.5}_{-0.5}$ & $8.6^{+0.4}_{-0.5}$ & $8.0^{+0.4}_{-0.4}$ & $7.7^{+0.4}_{-0.4}$ & $7.2^{+0.3}_{-0.4}$ \\
{\tt zxipcf} & $N_\mathrm{H}$ (10$^{23}$cm$^{-2}$; fixed) & 5.7 & 5.7 & 5.7 & 5.7 & 5.7 \\
 & log $\xi$ (fixed) & 2.7 & 2.7 & 2.7 & 2.7 & 2.7 \\
 & CF & $0.58^{+0.02}_{-0.02}$ & $0.67^{+0.01}_{-0.01}$ & $0.72^{+0.01}_{-0.01}$ & $0.76^{+0.01}_{-0.01}$ & $0.78^{+0.01}_{-0.01}$ \\
 & $v_\mathrm{out}$ (/c) & $-0.29^{+0.01}_{-0.01}$ & $-0.28^{+0.00}_{-0.00}$ & $-0.27^{+0.00}_{-0.00}$ & $-0.26^{+0.00}_{-0.00}$ & $-0.26^{+0.00}_{-0.00}$ \\ 
{\tt powerlaw} & $\Gamma$ & $2.78^{+0.02}_{-0.02}$ & $2.80^{+0.02}_{-0.02}$ & $2.77^{+0.02}_{-0.03}$ & $2.76^{+0.03}_{-0.03}$ & $2.69^{+0.03}_{-0.04}$ \\ 
 & norm (10$^{-3}$) & $5.0^{+0.2}_{-0.2}$ & $3.6^{+0.2}_{-0.1}$ & $2.9^{+0.1}_{-0.1}$ & $2.3^{+0.1}_{-0.1}$ & $1.4^{+0.1}_{-0.1}$ \\ 
{\tt diskbb} & $kT_{\rm in}$ (eV) & $0.24^{+0.01}_{-0.01}$ & $0.22^{+0.01}_{-0.01}$ & $0.21^{+0.01}_{-0.01}$ & $0.20^{+0.01}_{-0.00}$ & $0.18^{+0.00}_{-0.00}$ \\ 
 & norm (10$^{2}$) & $2.3^{+0.6}_{-0.5}$ & $3.1^{+0.5}_{-0.5}$ & $3.0^{+0.5}_{-0.5}$ & $3.4^{+0.6}_{-0.6}$ & $4.0^{+0.6}_{-0.5}$ \\ 
{\tt kabs} & $N_{\rm \ion{Fe}{XXV}}$ (10$^{18}$cm$^{-2}$) & $10^{+3}_{-2}$ & $7^{+6}_{-2}$ & $8^{+3}_{-2}$ & $15^{+8}_{-4}$ & $11^{+4}_{-7}$ \\ 
 & $z$ & $-0.15^{+0.01}_{-0.01}$ & $-0.15^{+0.01}_{-0.02}$ & $-0.15^{+0.01}_{-0.01}$ & $-0.15^{+0.01}_{-0.01}$ & $-0.14^{+0.03}_{-0.01}$ \\ 
{\tt kabs} & $N_{\rm \ion{Fe}{XXV}}$ (10$^{18}$cm$^{-2}$) & $13^{+9}_{-4}$ & $8^{+4}_{-3}$ & $14^{+11}_{-5}$ & $13^{+13}_{-6}$ & $6^{+3}_{-2}$ \\ 
 & $z$ & $-0.25^{+0.01}_{-0.01}$ & $-0.22^{+0.01}_{-0.02}$ & $-0.23^{+0.01}_{-0.01}$ & $-0.26^{+0.01}_{-0.01}$ & $-0.21^{+0.05}_{-0.01}$ \\ \hline
 $\chi^2$/dof & & 766.1/570 & 704.5/541 & 739.9/535 & 780.3/524 & 763.8/499 \\\hline
\end{tabular}
}
\end{table*}


\subsection{UV--X-ray SED fitting} \label{sec:4.3}
To investigate the outflow driving mechanism, we performed broadband Spectral Energy Distribution (SED) fitting using both X-ray (pn) and UV (OM) data. Since the flux variability in the OM data was found to be less than 10~\% throughout the observation period for all the three sources, we utilized the time-averaged OM spectrum for the analysis.

Given that PDS~456 and 1H~0707$-$495 are super-Eddington objects with significant AGN contributions, host galaxy contamination was considered negligible \citep{Castello-Mor16, Done16, Parker17}. 
For SED fitting, we assumed that the spectrum A in PDS 456 is free of partial absorption (CF $=0$).
Since the spectrum A in 1H 0707--495 is obscured, the SED model was multiplied by the {\tt zxipcf} model, where the parameters are fixed to those obtained by the spectral fitting (Tables \ref{tab:best_specfit_1H0707_2008} and \ref{tab:best_specfit_1H0707_2010}).
The SED is modeled by the {\tt agnslim} model \citep{Kubota19}, which describes the broadband emission of super-Eddington AGNs based on the slim disk model.
The {\tt agnslim} model assumes a radially stratified accretion flow with three key regions: 
\begin{itemize}
  \item An inner hot Comptonization region ($R_\mathrm{in}$ to $R_\mathrm{hot}$) with electron temperature $kT_\mathrm{hot}$ and photon index $\Gamma_\mathrm{hot}$, 
  \item An intermediate warm Comptonization region responsible for the soft X-ray excess ($R_\mathrm{warm}$ to $R_\mathrm{hot}$) with electron temperature $kT_\mathrm{warm}$ and photon index $\Gamma_\mathrm{warm}$, 
  \item An outer standard disk extending from $R_\mathrm{warm}$ to $R_\mathrm{out}$.
\end{itemize}

For highly super-Eddington slim disks, the inner disk radius ($R_\mathrm{in}$) may become
smaller than the innermost stable circular orbit, because it is determined not only by the black
hole spin but also by the gas pressure \citep{Watarai00, Kubota19}. In this study, we assumed a zero spin parameter ($a_*=0$) and set $R_\mathrm{in}$ according to the formula provided by \citet{Kubota19}. 
The interstellar absorption was fixed at $N_\mathrm{H} = 5 \times 10^{20}$ cm$^{-2}$ \citep{Bekhti16}, with the reddening parameter set to $E(B-V) = 1.7 \times N_\mathrm{H}/10^{22}$ cm$^{-2}$ \citep{Bohlin78}. 

\begin{figure}
\centerline{\includegraphics[width=1.0\columnwidth]{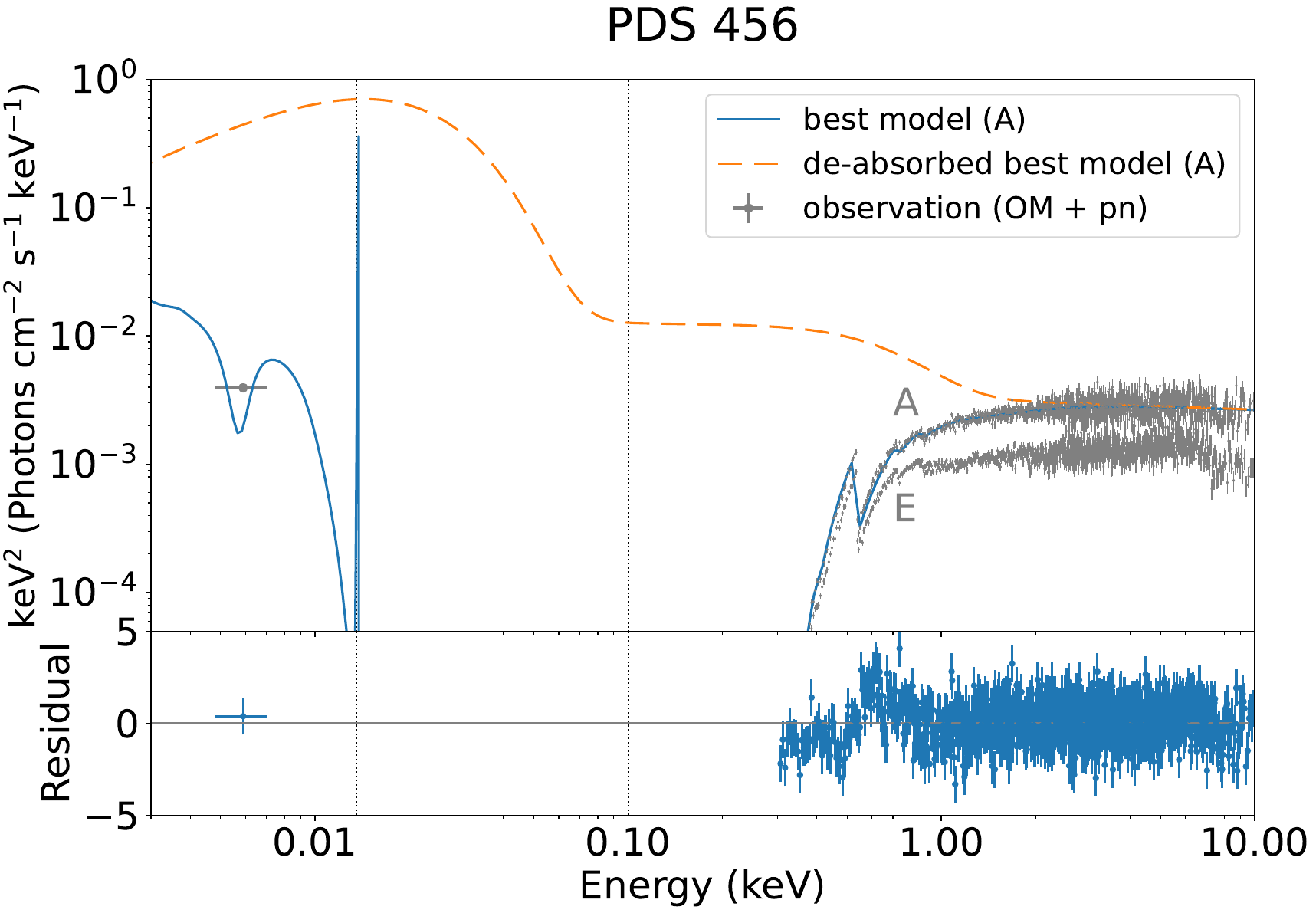}} 
\centerline{\includegraphics[width=1.0\columnwidth]{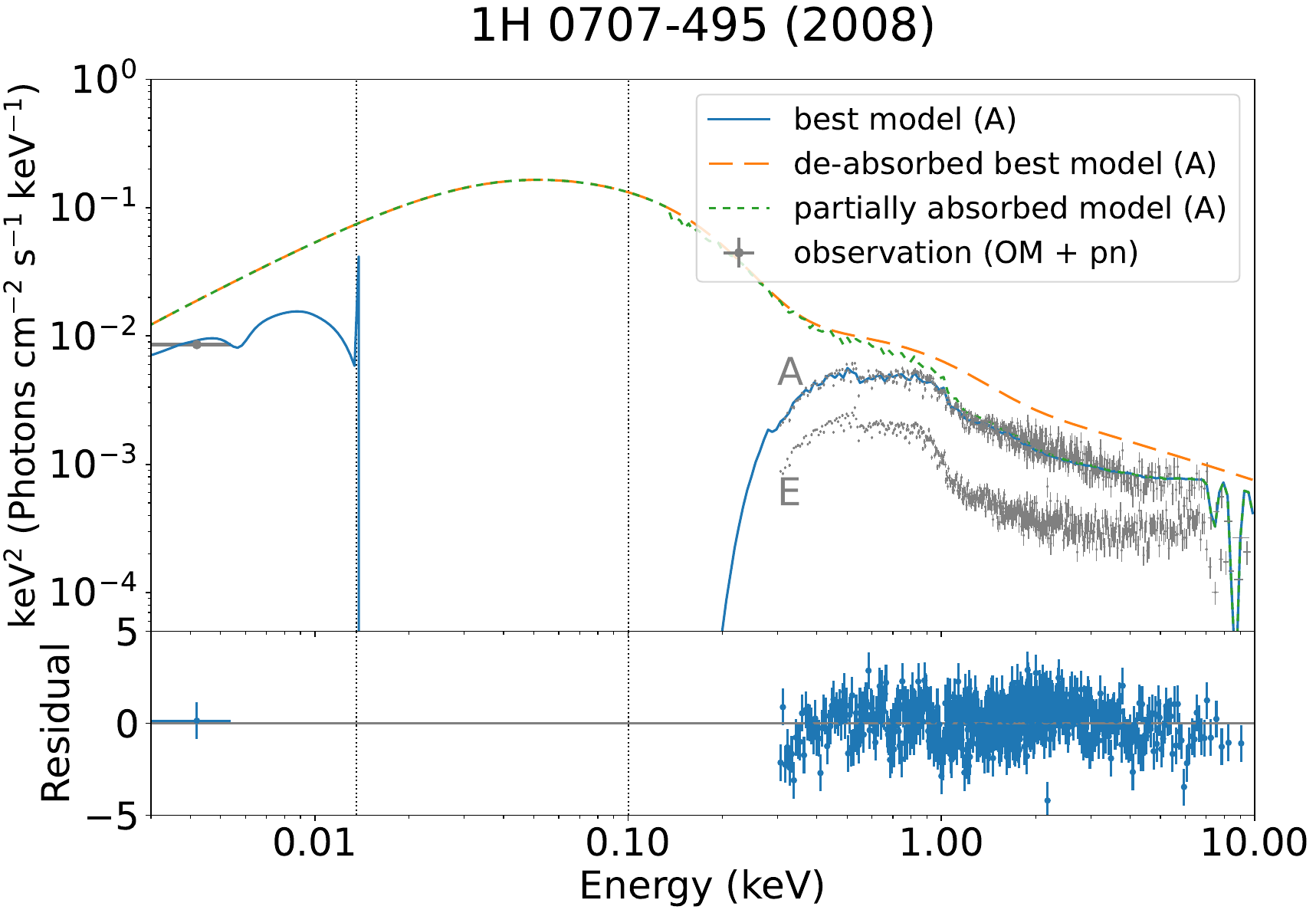}} 
\centerline{\includegraphics[width=1.0\columnwidth]{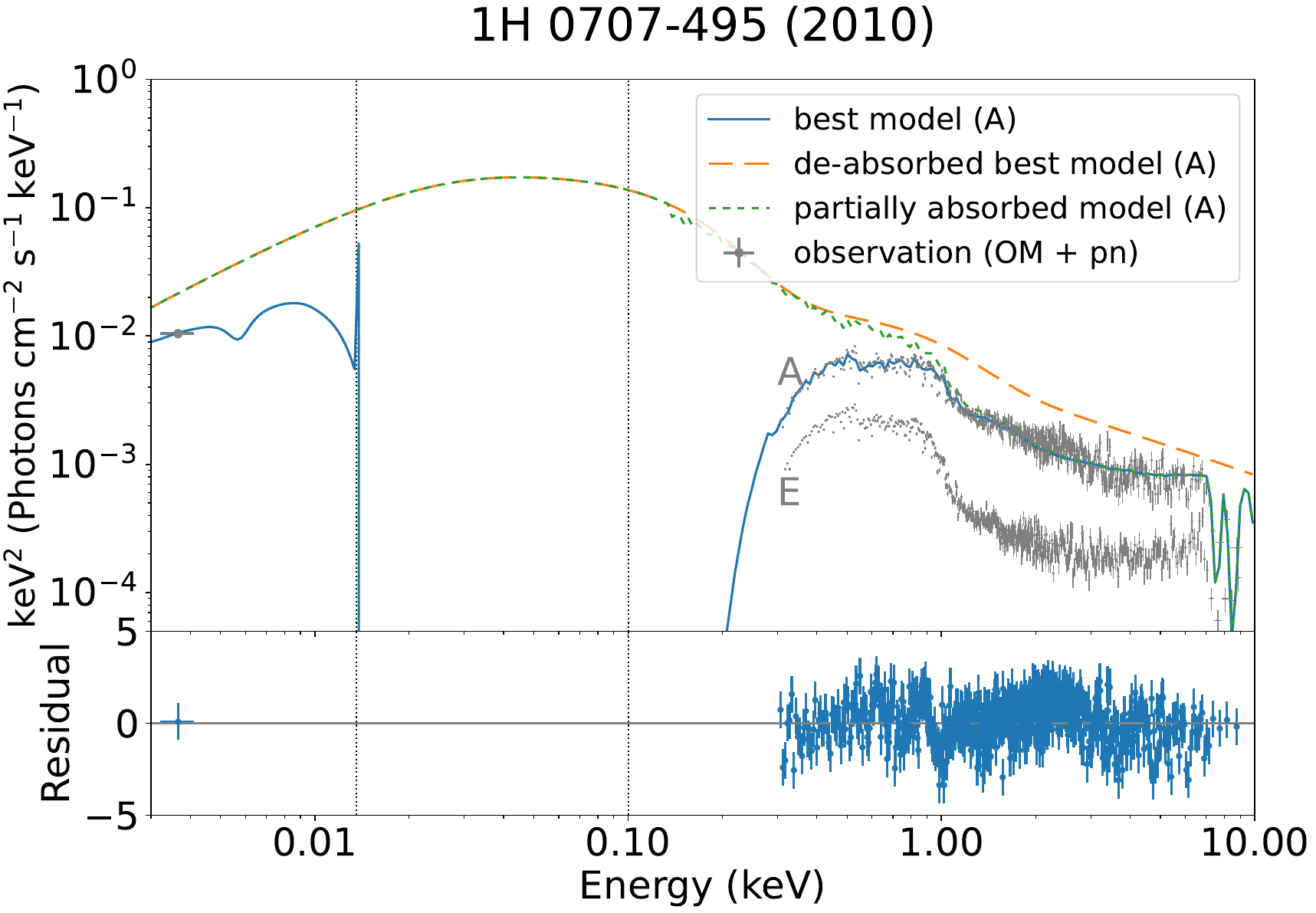}} 
	 \caption{Broadband UV--X-ray SED fitting of PDS 456 and 1H 0707--495 (2008, 2010) with pn and OM data in 3~eV--10~keV. The two gray lines show the intensity-sliced spectra A and E. The blue line shows the best-fit model for the spectrum A. The orange dashed line shows the de-absorbed best-fit model. The vertical dotted lines show the boundary of the energy bands, 13.6~eV and 0.1~keV.\\
{Alt text: There are 3 figures. The upper panel shows the $\nu F_\nu$ plot with the best fit model. The lowe panel shows (data--model)/error.}} 
	 \label{fig:agnslim}
\end{figure}

The best-fit SED model for the spectrum A is presented in Figure~\ref{fig:agnslim}, showing the broadband emission in the 3 eV--10 keV range. The derived fit parameters are summarized in Table~\ref{tab:agnslim_PDS456}--\ref{tab:agnslim_1H0707_2010}. The results suggest that the AGN emission is dominated by the disk component in the UV band and by the Comptonized emission in the X-ray band.

\begin{table}
\centering
\caption{Best-fit parameters determined by the SED fitting of PDS 456 obtained with the OM and pn instruments.}
 \label{tab:agnslim_PDS456}
\scalebox{1.0}{
 \begin{tabular}{llc}\hline
 Components & Parameters & Best-fit values \\\hline
 {\tt phabs} & $N_\mathrm{H}$ (10$^{21}$cm$^{-2}$) & 3.7 (fixed) \\
 {\tt redden} & $E(B-V)$ & 0.63 (fixed) \\
 {\tt agnslim} & mass ($M_\odot$) & 1.0$\times 10^9$ (fixed) \\
 & $\dot{m}$~($=\dot{M}/\dot{M}_{\rm Edd}$) & 2.0$\pm0.7$ \\
 & $a^*$ & 0 (fixed) \\
 & cos~$i$ & 0.5 (fixed) \\
 & $kT_{\rm hot}$ (keV) & 100 (fixed) \\
 & $kT_{\rm warm}$ (keV) & 0.156$\pm0.005$ \\
 & $\Gamma_{\rm hot}$ & 2.09$\pm0.02$ \\
 & $\Gamma_{\rm warm}$ & 1.6$\pm0.3$ \\
 & $R_{\rm hot}$ (R$_{\rm g}$) & 7.9$^{+0.3}_{-0.4}$ \\
 & $R_{\rm warm}$ (R$_{\rm g}$) & 8.4$^{+0.2}_{-0.5}$ \\\hline
	\end{tabular}}
\end{table}
\begin{table}
\centering
 \caption{The same as Table~\ref{tab:agnslim_PDS456}, but for 1H~0707$-$495 in 2008.}
 \label{tab:agnslim_1H0707_2008}
\scalebox{1.0}{
 \begin{tabular}{llc}\hline
 Components & Parameters & Best-fit values \\\hline
 {\tt phabs} & $N_\mathrm{H}$ (10$^{20}$cm$^{-2}$) & 8.0 (fixed) \\
 {\tt redden} & $E(B-V)$ & 0.136 (fixed) \\
{\tt zxipcf} & $N_\mathrm{H}$ (10$^{23}$cm$^{-2}$) & 5.6 (fixed) \\
 & log $\xi$ & 2.65 (fixed) \\
 & CF & 0.52 (fixed) \\
 & $v_\mathrm{out}$ (/c) & $-$0.29 (fixed)\\ 
 {\tt agnslim} & mass ($M_\odot$) & 4.0$\times 10^6$ (fixed) \\
 & $\dot{m}$~($=\dot{M}/\dot{M}_{\rm Edd}$) & 9.6$\pm0.1$ \\
 & $a^*$ & 0 (fixed) \\
 & cos~$i$ & 0.5 (fixed) \\
 & $kT_{\rm hot}$ (keV) & 100 (fixed) \\
 & $kT_{\rm warm}$ (keV) & 0.20$\pm0.01$ \\
 & $\Gamma_{\rm hot}$ & 2.71$\pm0.02$ \\
 & $\Gamma_{\rm warm}$ & 1.60$\pm{0.3}$ \\
 & $R_{\rm hot}$ (R$_{\rm g}$) & 6.5$\pm0.1$ \\
 & $R_{\rm warm}$ (R$_{\rm g}$) & 7.0$\pm0.1$ \\\hline
	\end{tabular}}
\end{table}
\begin{table}
\centering
 \caption{The same as Table~\ref{tab:agnslim_PDS456}, but for 1H~0707$-$495 in 2010.}
 \label{tab:agnslim_1H0707_2010}
\scalebox{1.0}{
 \begin{tabular}{llc}\hline
 Components & Parameters & Best-fit values \\\hline
 {\tt phabs} & $N_\mathrm{H}$ (10$^{20}$cm$^{-2}$) & 9.0 (fixed) \\
 {\tt redden} & $E(B-V)$ & 0.153 (fixed) \\
{\tt zxipcf} & $N_\mathrm{H}$ (10$^{23}$cm$^{-2}$) & 5.6 (fixed) \\
 & log $\xi$ & 2.73 (fixed) \\
 & CF & 0.58 (fixed) \\
 & $v_\mathrm{out}$ (/c) & $-$0.29 (fixed)\\ 
 {\tt agnslim} & mass ($M_\odot$) & 4.0$\times 10^6$ (fixed) \\
 & $\dot{m}$~($=\dot{M}/\dot{M}_{\rm Edd}$) & 15.2$\pm0.1$ \\
 & $a^*$ & 0 (fixed) \\
 & cos~$i$ & 0.5 (fixed) \\
 & $kT_{\rm hot}$ (keV) & 100 (fixed) \\
 & $kT_{\rm warm}$ (keV) & 0.20$\pm0.01$ \\
 & $\Gamma_{\rm hot}$ & 2.77$\pm0.02$ \\
 & $\Gamma_{\rm warm}$ & 1.60$\pm{0.2}$ \\
 & $R_{\rm hot}$ (R$_{\rm g}$) & 5.8$\pm0.1$ \\
 & $R_{\rm warm}$ (R$_{\rm g}$) & 6.4$\pm0.1$ \\\hline
	\end{tabular}}
\end{table}

\section{Discussion} \label{Chapter5} 
\subsection{Effectiveness of the spectral-ratio model fitting}\label{sec:5.1}
The spectra-ratio model fitting method developed by \citet{Midooka22a} was systematically applied to several Seyfert~1 galaxies, and we have found an evidence of the ultra-fast clumpy soft X-ray absorbers in PDS~456, 1H~0707$-$495, and Mrk~335, very similar to the one in IRAS 13224--3809 \citep{midooka23}. 
\citep{midooka23}.
While such an ultra-fast soft X-ray wind has been reported in PDS 456 \citep{reeves2020b}, this is the first report in other Seyfert galaxies.
Here, we note that there are certain conditions under which the spectral-ratio fitting method becomes effective in constraining the clump velocities.
First, the spectral-ratio fitting requires several intensity-sliced spectra with enough photon statistics. The X-ray count rates in 0.3--10~keV of roughly at least 1~counts~s$^{-1}$ are  required 
even when the source is the dimmest; among the 12 sources we studied, 
the X-ray fluxes of ESO~323--G77 and PG~1126--041 were too low to apply the spectral-ratio fitting method
 (Table~\ref{tab:obs_log}).
In addition, the spectral ratios need to represent characteristic spectral variations, so that sufficient time variation is required both in the X-ray flux and the spectral shape. For NGC~985 (2013), NGC~7314, and Ton~S180, the spectral variability was so small that little information was extracted by taking the spectral ratios.

Main contributors to the X-ray spectral-ratio are variations of the intrinsic X-ray luminosity and the X-ray absorber, but the former does not cause energy dependence.
Thus, the key to disentangling the parameter degeneracy is the X-ray absorber variation, which is manifested in the characteristic dip/cliff structure in the spectral-ratio of PDS~456, 1H~0707$-$495 and Mrk~335 (Figures~\ref{fig:eachfit}), as well as IRAS 13224--3809 \citep{midooka23}.
The appearance of the dip/cliff structure in the spectral-ratio is essential to estimate reliable
absorber parameters in this method.
In fact, in the ratio fitting of Mrk~766 and NGC~985,
the absorber parameters were hardly constrained, since the spectral ratios were almost flat. 

The clumpy absorber have three parameters, $N_\mathrm{H}$, $\xi$, and $v_\mathrm{out}$.
We found that $N_\mathrm{H}$ and $\xi$ are invariable for different flux levels, 
thus the dip structure appears when there are 
significant changes in $v_\mathrm{out}$.
The energy shift of the dip gives us the clump velocity in the line-of-sight, which makes this method unique to directly constrain the radial velocity of the partial absorbing clouds.
Finally, we remark that taking the ratio can cancel out the ``multiplicative'' contributions such as constant absorption, while it cannot cancel out the ``additive'' contributions such as the constant additional emission component seen in NGC~3783.


\subsection{The outflowing velocities of the UFO and the clumpy absorbers}\label{sec:5.2} 
In section~\ref{sec:4.1}, we constrained the outflow velocity $v_\mathrm{out}$ of the partial clumpy absorbers using the spectral-ratio fitting method. We also performed conventional spectral fitting to determine the UFO velocity $v_\mathrm{UFO}$ in section~\ref{sec:4.2}.

\begin{figure*}
 \begin{minipage}{0.99\columnwidth}
 \centering
\includegraphics[width=0.95\columnwidth]{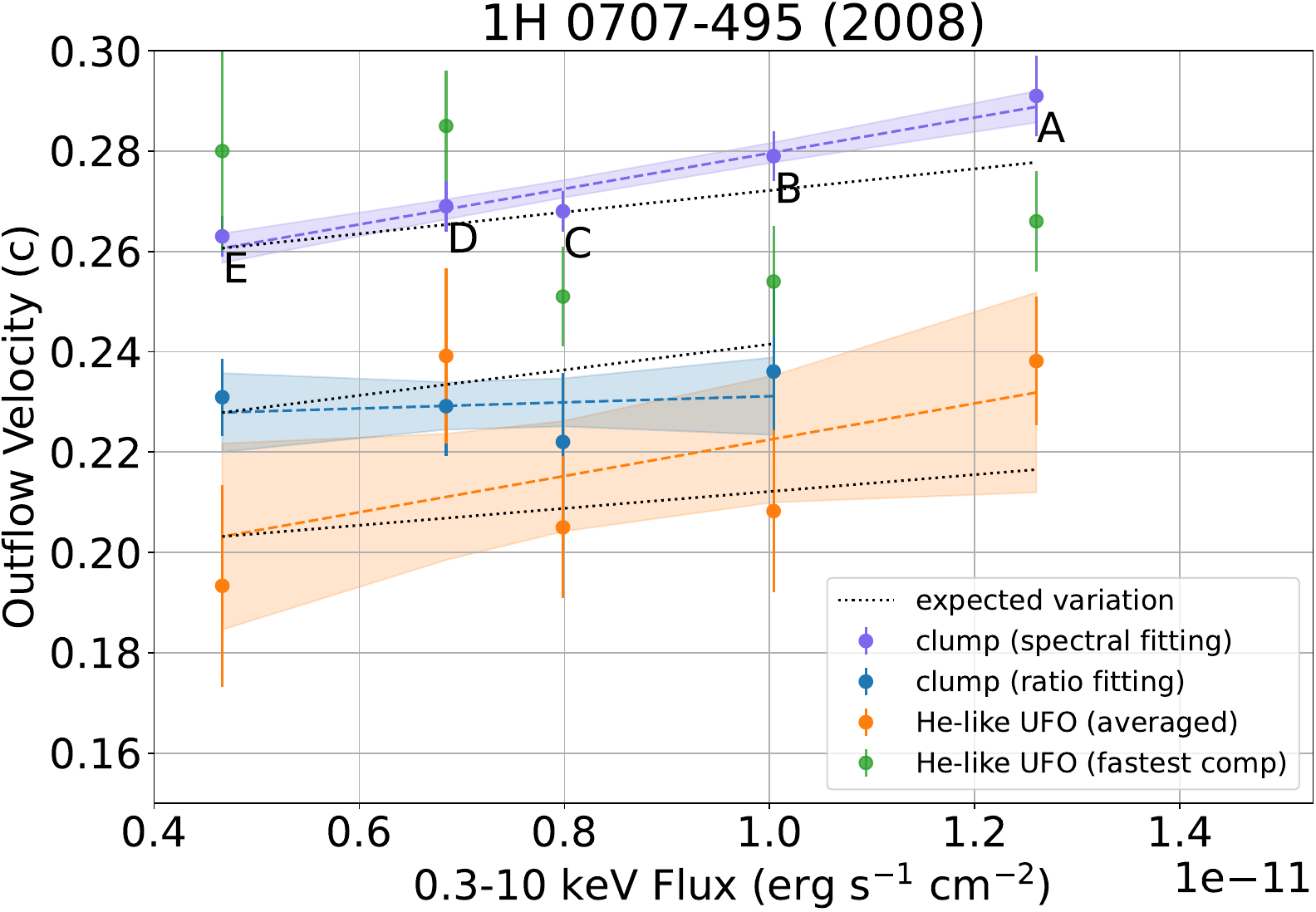}
  \end{minipage} 
 \begin{minipage}{0.99\columnwidth}
\centering
\includegraphics[width=0.95\columnwidth]{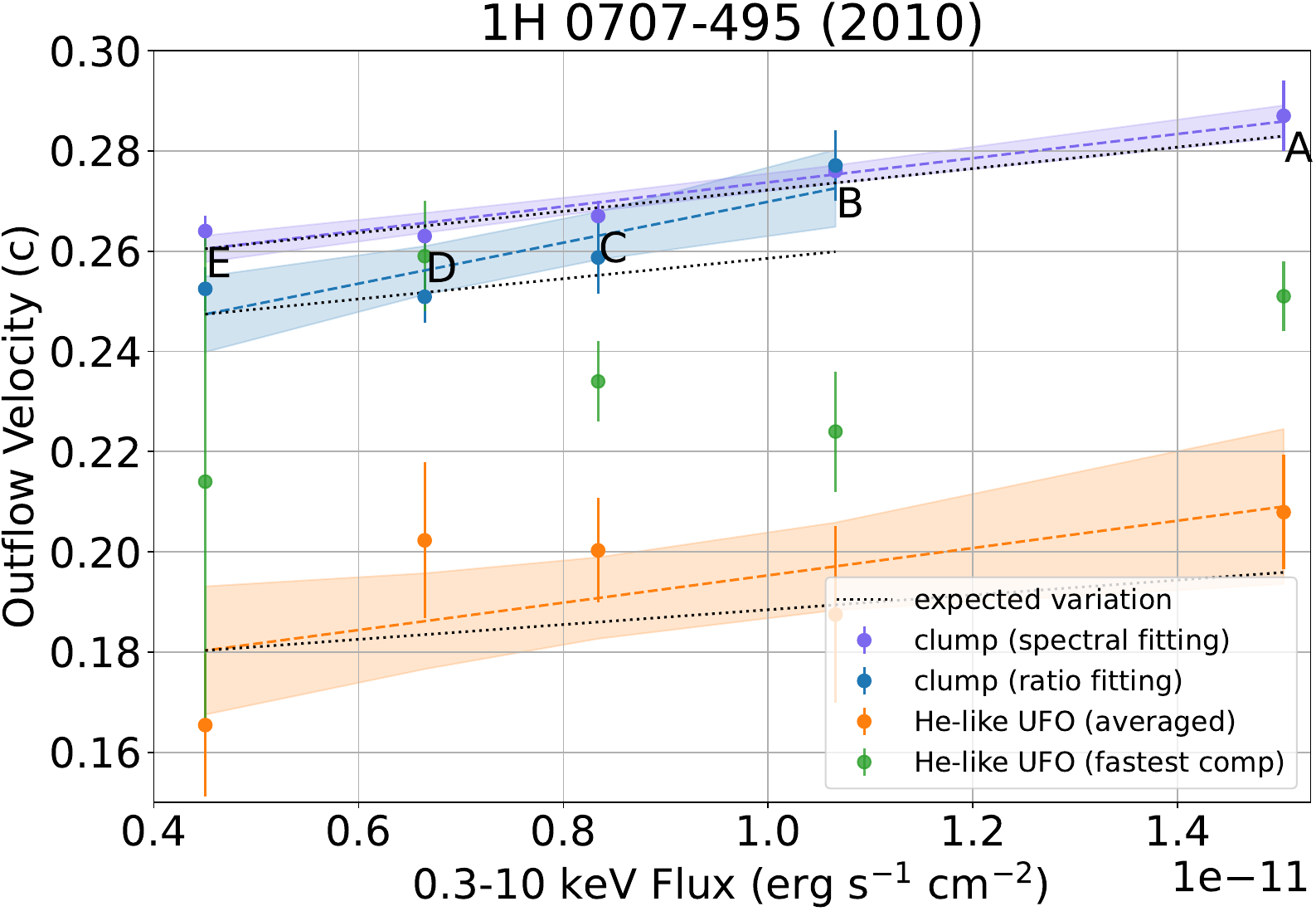}
  \end{minipage} \\
  \vspace{3mm}
 \begin{minipage}{0.99\columnwidth}
 \centering
\includegraphics[width=0.95\columnwidth]{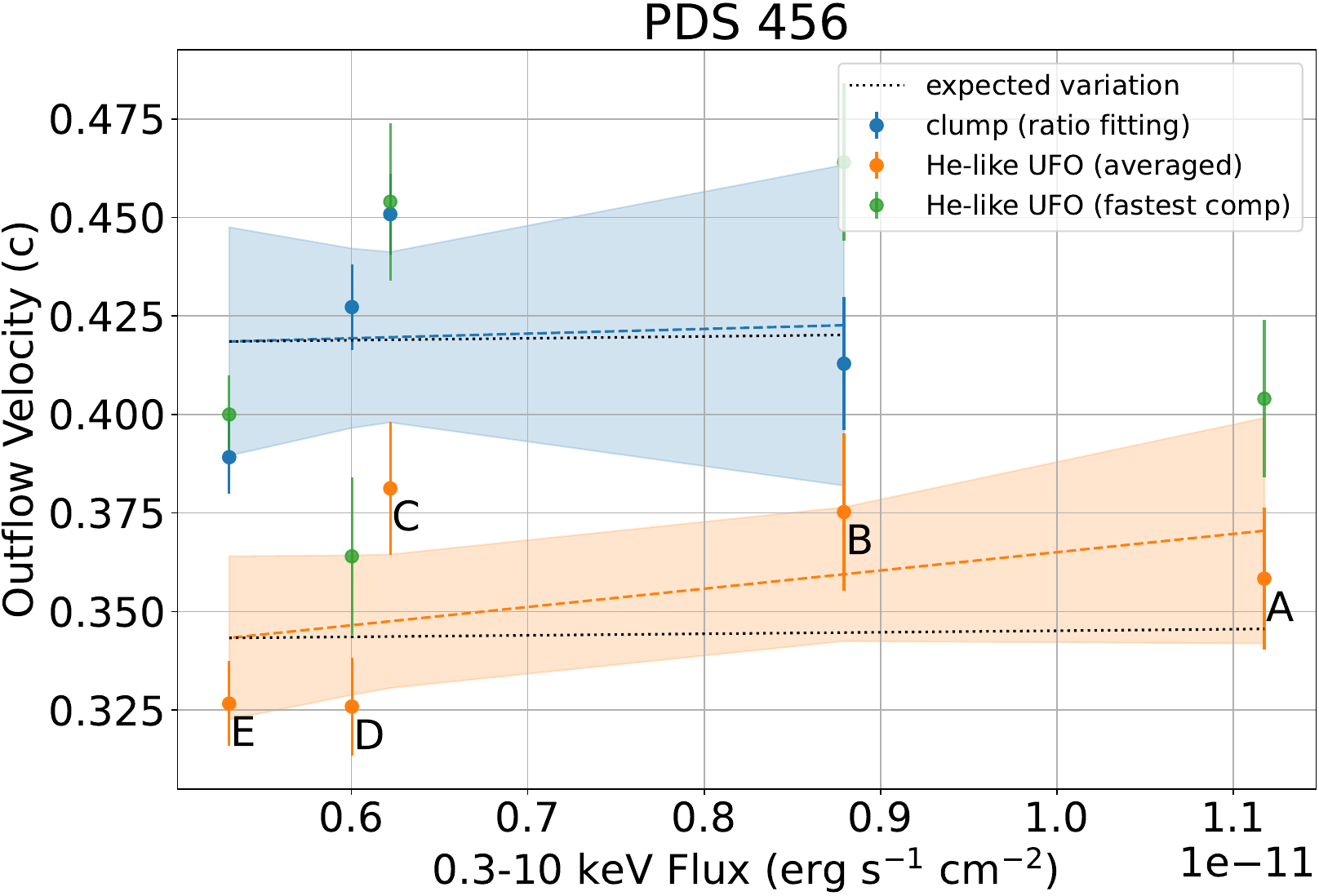}
  \end{minipage} 
 \begin{minipage}{0.99\columnwidth}
 \centering
\includegraphics[width=0.95\columnwidth]{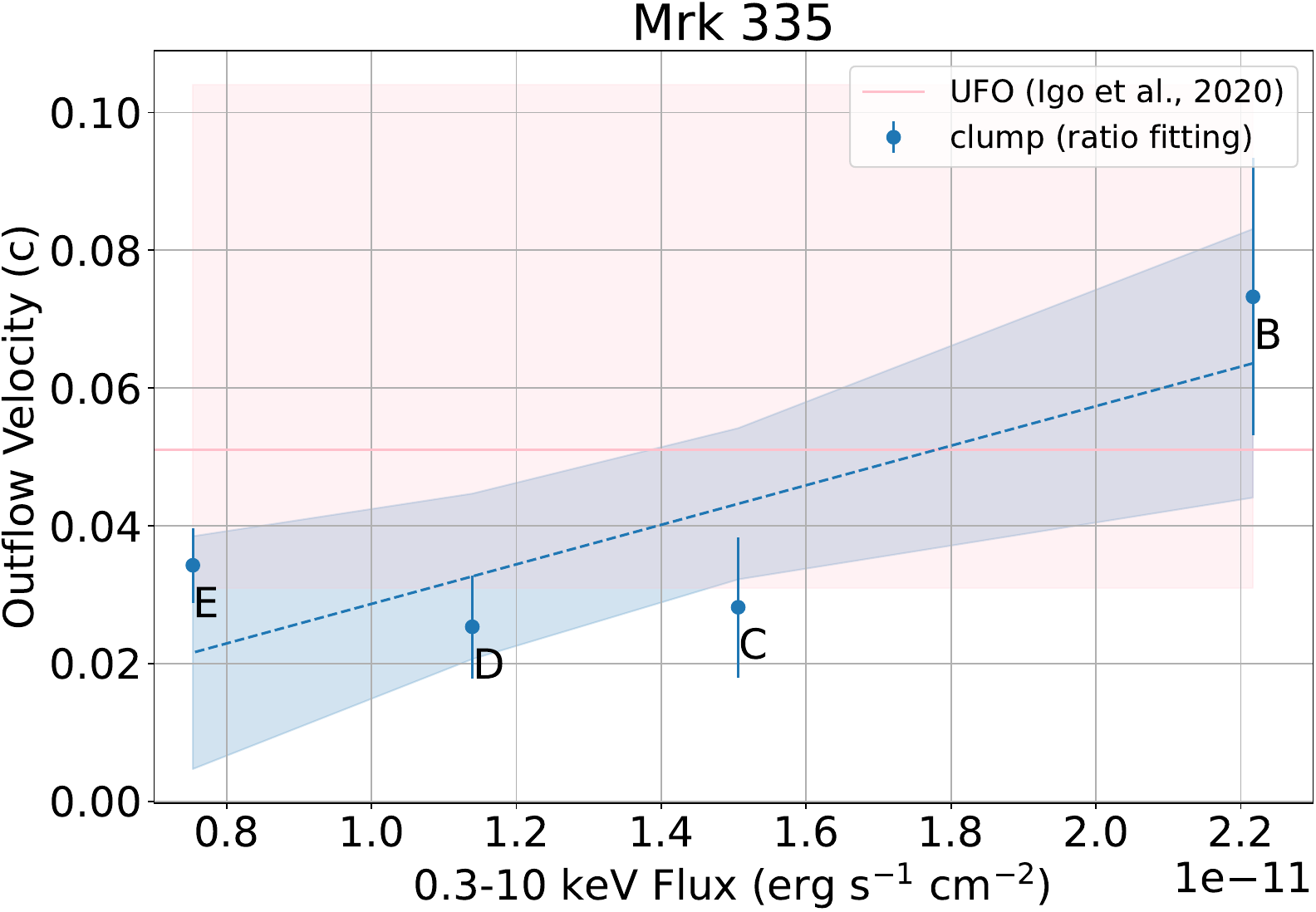}
  \end{minipage} 
  \vspace{3mm}
\caption{  
Comparison of the velocities between the UFOs and clumps. Outflow velocities of the clumpy absorbers obtained from the spectral-ratio fitting (in blue), those from the spectral fitting (in purple, only for 1H 0707--495), and those of the UFOs estimated from the spectral fitting (in orange). The outflowing velocities are plotted against the 0.3--10~keV flux of the intensity-sliced spectra. The blue, purple, and orange dotted lines show the best-fit linear functions of each with 1$\sigma$ uncertainties. The fastest UFO component obtained with spectral fitting (in green) and expected velocity variations are also plotted (in black). In Mrk 335, the UFO velocity in \citet{Igo20} is plotted with its uncertainty in pink.
\\
{Alt text: There are 4 figures. }
}
\label{fig:velcomp}
  \end{figure*}

Comparison of the outflow velocities of the UFOs and the clumpy absorbers is shown in Figures~\ref{fig:velcomp}, where these velocities are plotted against the 0.3--10.0~keV X-ray flux of the intensity-sliced spectra.
The clump velocities obtained by both the ratio fitting and the spectral fitting are shown for 1H~0707$-$495.
For PDS~456, the clump velocity in the spectral fitting was fixed to the value obtained by the spectral-ratio fitting. 
When the UFO feature is composed of multiple absorption lines with different velocity components, the average velocity weighted by the column density of each absorption line ($N_{\rm ion}$) is plotted (see Tables~\ref{tab:best_specfit_PDS456}--\ref{tab:best_specfit_1H0707_2010}). 

We fitted the velocities of the UFOs and clumpy absorbers as a function of the X-ray flux with linear functions.
The blue, purple, and orange dotted lines and areas in Figure~\ref{fig:velcomp} show the best fit models and their 1~$\sigma$ uncertainties,
respectively.
Since the UFO absorption of Mrk~335 in our dataset is too tiny to estimate the velocity, the UFO velocity presented in the previous study (\cite{Igo20}; see Table~\ref{tab:obs_target}) is plotted with its uncertainty in pink. Note that \citet{Igo20} did not perform the UFO search for flux resolved spectra, so the quoted UFO velocity is common to all the X-ray fluxes.
Although the present analysis indicates the high UFO and clump velocities ($>$0.1 $c$) for 1H~0707$-$495 and PDS~456, the clump velocity of Mrk 335 is much smaller ($<$0.1 $c$) in agreement with the small UFO velocities suggested in the previous works \citep{Gallo19, Igo20}. 
These  agreements between the UFO velocities and the clump velocities
strengthen the reliability of the estimation of the clump velocity with the spectral-ratio model fitting.

We note that there is a systematic offset between the clump velocities derived from the spectral-ratio fitting (blue points) and the direct spectral fitting (purple points) in 1H 0707--495. This discrepancy is likely attributable to the difference in the model assumptions. In the spectral-ratio fitting, the brightest spectrum (A) is assumed to be unobscured ($\mathrm{CF}=0$) to serve as a baseline. However, the direct spectral fitting reveals that Spectrum A still retains a significant covering fraction of $\mathrm{CF} \simeq 0.5$ (see Tables \ref{tab:best_specfit_1H0707_2008}, \ref{tab:best_specfit_1H0707_2010}). This non-zero absorption in the denominator of the spectral ratios introduces a systematic bias in the derived parameters. Additionally, the difference in the energy band used (0.3--5.0 keV for ratio fitting vs 0.3--10.0 keV for spectral fitting) and the sensitivity to the soft excess modeling may also contribute to the parameter offsets. Nevertheless, it is important to emphasize that both independent methods consistently indicate UFO velocities ($v_\mathrm{out} > 0.1c$) that are positively correlated with the source flux. This robustly supports the conclusion that the clumpy absorbers are kinematically linked to the UFOs.

To quantify the trends shown in Figure \ref{fig:velcomp}, we evaluated the correlation between the outflow velocity and the X-ray flux. For Mrk 335 and 1H 0707--495 (2010), the Pearson correlation coefficients are derived to be $r \simeq 0.86$ and $r \simeq 0.94$, respectively, indicating a strong positive correlation. On the other hand, 1H 0707--495 (2008) and PDS 456 show weaker correlation ($r \simeq 0.22$ and 0.40, respectively), possibly due to the larger statistical uncertainties or intrinsic variability during that epoch. However, considering the consistent positive slopes observed in the most significant cases (and also in the UFO velocity components), we interpret the overall results as supporting the scenario in which the outflow kinetic energy increases with the intrinsic luminosity. This observational connection serves as the premise for the discussion on the driving mechanism in the following Section \ref{sec4.3}.

\subsection{Possibility of the line-force acceleration of the clumpy absorbers}\label{sec4.3}
The outflowing clump, as well as the UFO, are considered to be driven by the radiation pressure, since their velocity is positively correlated with the X-ray flux, as discussed in the previous subsection.
It is also seen that the clump velocities are often faster than the UFO velocity components (Figure \ref{fig:velcomp}). 
This suggests that some additional mechanisms may be working only for the clumps.
Here, we propose a hypothesis that the clumps are accelerated by the line force in addition to the continuum force.
It is considered that the line-force-driven outflow is caused by a mildly ionized gas, which efficiently absorbs the UV radiation in the SED.
Since the mildly ionized gas is commonly located at the accretion disk in the {\em sub-}Eddington objects, it has not been considered, so far, to be a major cause of the outflow in the {\em super-}Eddington objects. Now, we suggest that the line force may also be effective in the super-Eddington objects,
considering that the emission weakens and the ionization parameter decreases as going away from the inner part of the disk.

We used the photoionization code XSTAR (version 2.57; \citealt{Kallman01, Kallman04}) to generate synthetic spectra
 affected by the mildly ionized gas.
The models describe transmission by the ionized absorbers as a function of the ionization parameter $\xi$, with the column density $N_\mathrm{H}$ fixed at $6\times 10^{23}$~cm$^{-2}$ (Tables \ref{tab:best_specfit_PDS456}--\ref{tab:best_specfit_1H0707_2010}).
We fixed the gas density at $1.0\times 10^{12}$~cm$^{-3}$, the gas temperature at $10^5$~K, and the turbulent velocity at 1,000~km~s$^{-1}$.
Abundances were fixed to the solar values except for those minor elements (Li, Be, Ba, Na, Al, P, Cl, K, Sc, Ti, V, Cr, Mn, Co, Cu, and Zn) that were ignored to simplify the calculation.
The incident spectrum was assumed to be the de-absorbed model with the X-ray flux in 13.6~eV--13.6~keV of $2.45 \times 10^{-10}$~erg~cm$^{-2}$~s$^{-1}$.

\begin{figure} 
\centering
\centerline{\includegraphics[width=0.99\columnwidth]{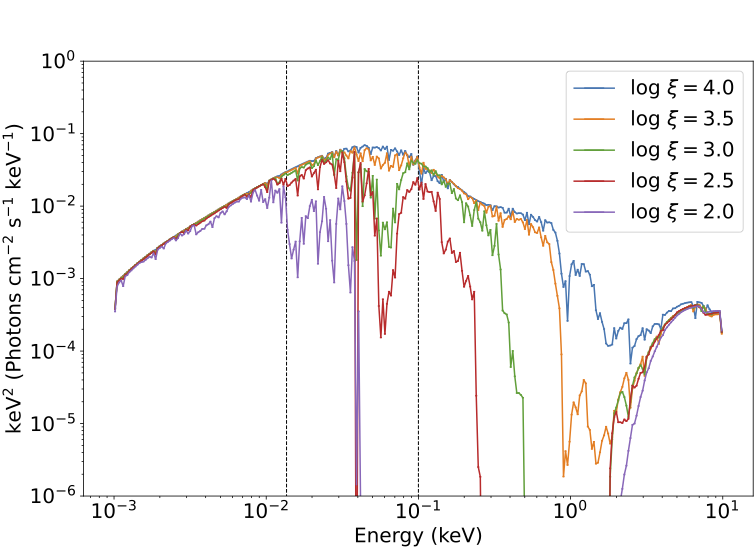}}
 \caption{The transmission spectra absorbed by ionized gas are shown, calculated with XSTAR for various log $\xi$ parameters.
 \\
{Alt text: The horizontal axis shows the X-ray energy from $10^{-3}$ to $10^{-1}$ keV. The vertical axis shows the $\nu F_\nu$ plot.}
} 
	 \label{fig:xstar_abscomp}
\end{figure}

Figure~\ref{fig:xstar_abscomp} shows the transmission spectra with input SEDs, where the interstellar absorption is de-absorbed.
In these spectra, log $\xi$ is varied at several different values from 2.0 to 4.0.
The input SED is heavily absorbed as the ionization parameter gets lower.
In addition to the characteristic dip structure around 1~keV, many UV absorption lines are found below 0.1~keV.

\begin{figure}
\centering
\centerline{\includegraphics[width=0.99\columnwidth]{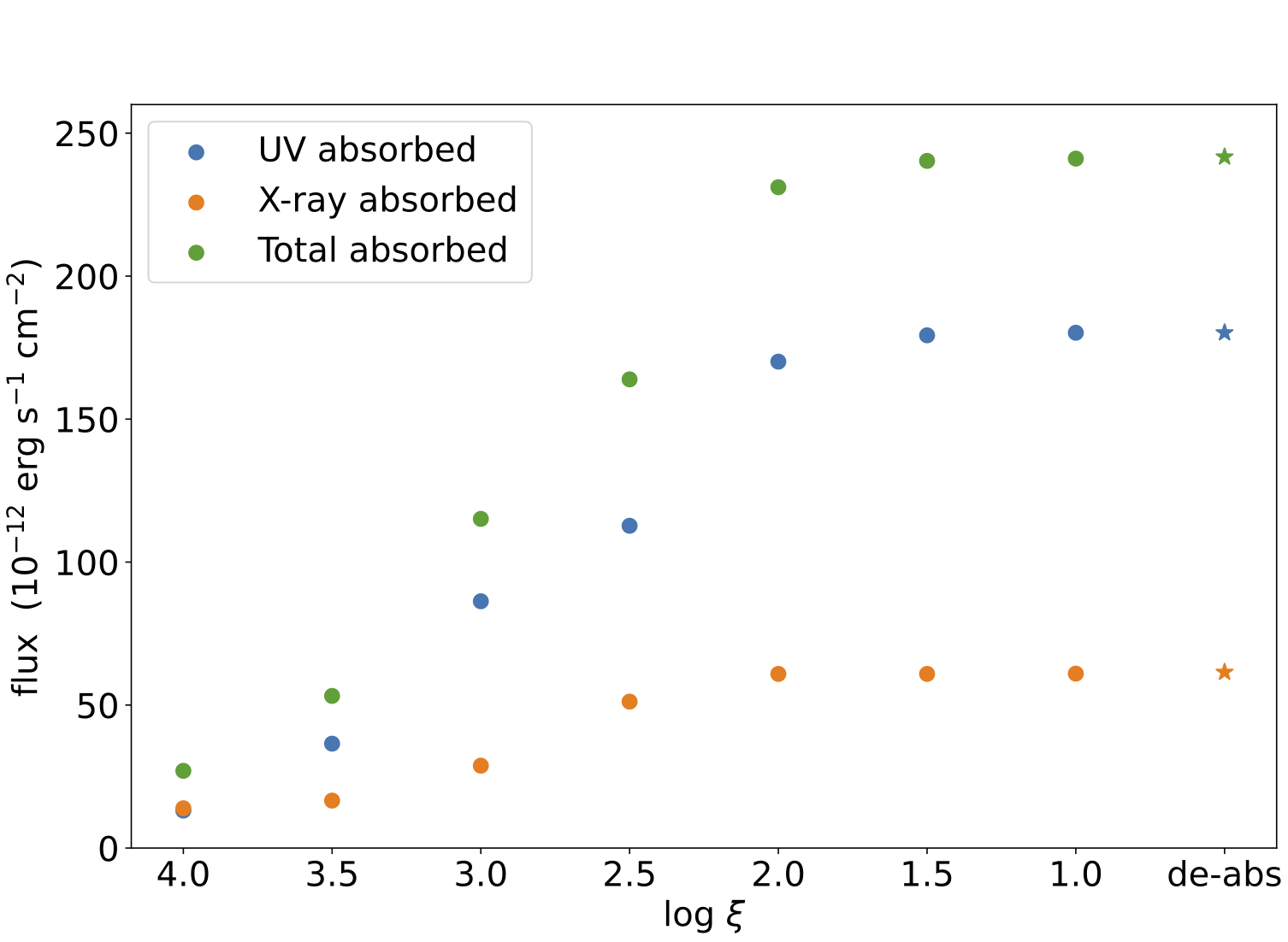}}
 \caption{Absorbed flux estimated with XSTAR. The blue, orange green dotted show the UV, X-ray, and total (UV + X-ray) absorbed fluxes. For comparison, the de-absorbed flux is plotted with the star marker. \\
{Alt text: The horizontal axis shows $\log\xi$. The vertical axis shows the flux ($10^{-12}$~erg~s$^{-1}$~cm$^{-2}$).}} 
	 \label{fig:absflux_comp}
\end{figure}

To estimate the line force contribution to the clump acceleration, we calculate how much the intrinsic UV and X-rays are intervened by the ionized gas, giving momentum to the gas.
The momentum received by the intervening cloud is estimated by dividing the absorbed energy flux by the speed of light.
The calculated UV and X-ray absorption fluxes for each ionization parameter are summarized in Figure~\ref{fig:absflux_comp}.
To quantitatively assess the contribution of the line force versus the continuum force, we compared the absorbed fluxes in Figure \ref{fig:absflux_comp} with the momentum transfer expected from Thomson scattering. The optical depth for Thomson scattering is calculated as $\tau_{\rm T} = \sigma_{\rm T} N_{\rm H} \simeq 0.37$ for the adopted column density of $N_{\rm H} = 6 \times 10^{23} \rm ~cm^{-2}$. 
With the intrinsic flux of $2.45 \times 10^{-10}$~erg~cm$^{-2}$~s$^{-1}$, the absorbed flux due to Thomson scattering is $0.9\times 10^{-10}$~erg~cm$^{-2}$~s$^{-1}$.
This continuum interaction provides a baseline momentum to the outflow. As shown in Figure \ref{fig:absflux_comp}, the absorbed flux for the moderately ionized clumps ($\log \xi \simeq 2.5$; Tables \ref{tab:best_specfit_PDS456} to \ref{tab:best_specfit_1H0707_2010}) is as large as $\sim1.6\times 10^{-10}$~erg~cm$^{-2}$~s$^{-1}$. This substantial increase in opacity indicates that the line force acts as a force multiplier, providing the additional acceleration required to drive the clumps to velocities comparable to or even exceeding those of the continuum-driven UFOs.

\begin{figure} 
\centerline{\includegraphics[width=1.0\columnwidth]{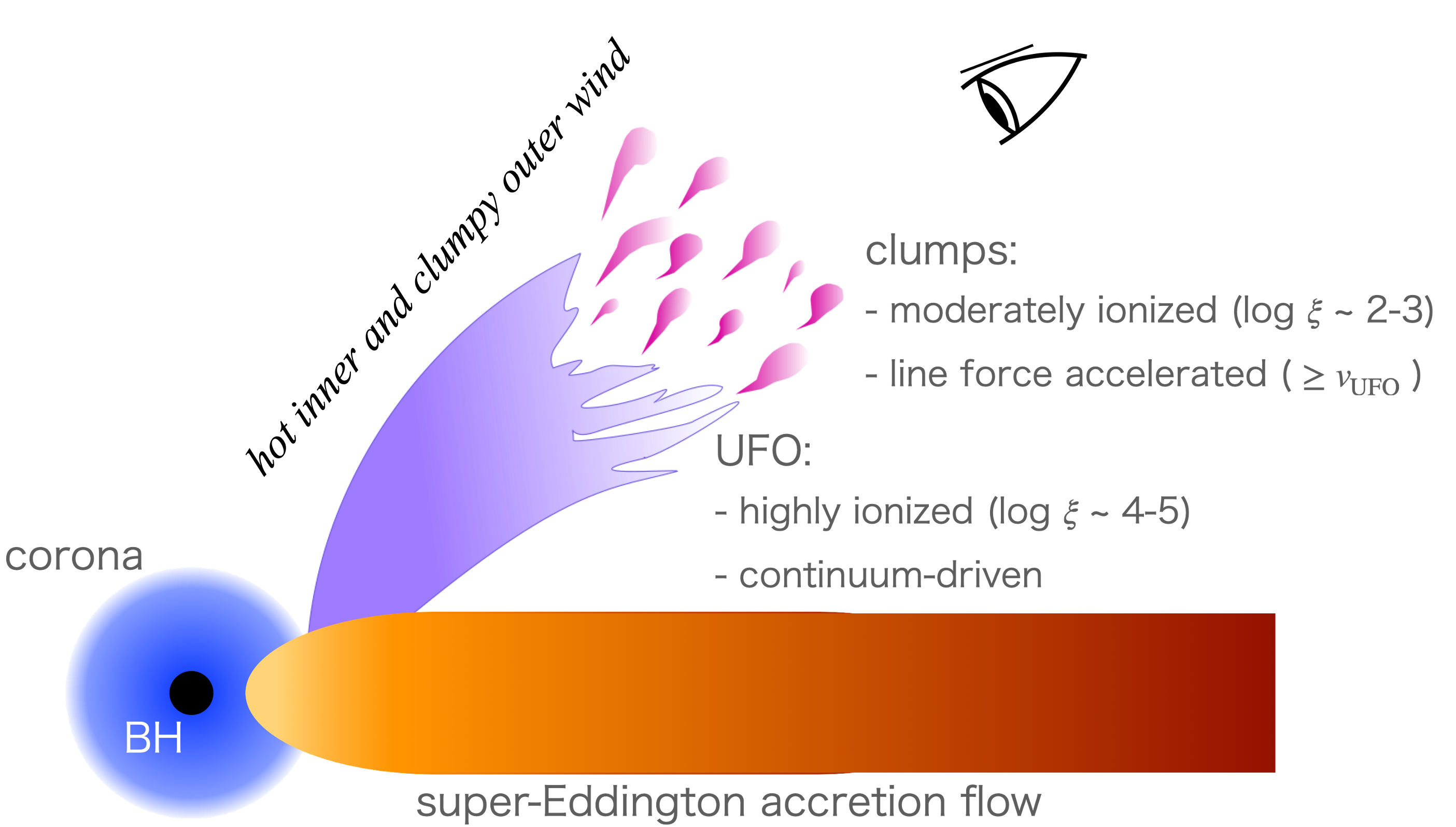}}
	 \caption{A schematic picture around a SMBH with super-Eddington accretion flow.\\
   {Alt text: The black hole with corona. the super-Eddington accretion flow, and hot inner and less-ionized clumpy outer wind are schematically shown.}
   }
	 \label{fig:sche_UFOclump}
\end{figure}

Figure~\ref{fig:sche_UFOclump} presents a schematic picture of the proposed environment surrounding the central SMBH of a super-Eddington object. 
The highly-ionized UFO is driven by the strong continuum radiation from the inner part of the super-Eddington accretion flow, as suggested by radiation hydrodynamical simulations (e.g., \citealt{Takeuchi13}, \citealt{Kitaki21}).
The wind turns into moderately ionized clumps away from the BH and is further accelerated by the line force.
Consequently, the ionized clump velocities can become higher than the UFO velocities.

\subsection{The unexplained $\sim$1 keV spectral feature} 
\label{sec:5.3}
1H~0707$-$495 (e.g., \citealt{Fabian09}) and 
IRAS~13224$-$3809 (e.g., \citealt{Ponti10})
have been known to exhibit unexplained characteristic spectral features at $\sim$1 keV. 
When partial covering models with {\em static}\/ absorbers were applied to these energy spectra, significant residuals were found at $\sim$1 keV, and unexplained strong absorption edges were additionally required to fit \citep{Mizumoto14, Yamasaki16}. 
Alternatively, an extremely strong disk reflection component was proposed to explain this $\sim$1 keV feature (e.g., \citealt{Parker17,Parker20,Pinto18}).

\begin{figure} 
\centerline{\includegraphics[width=1.0\columnwidth]{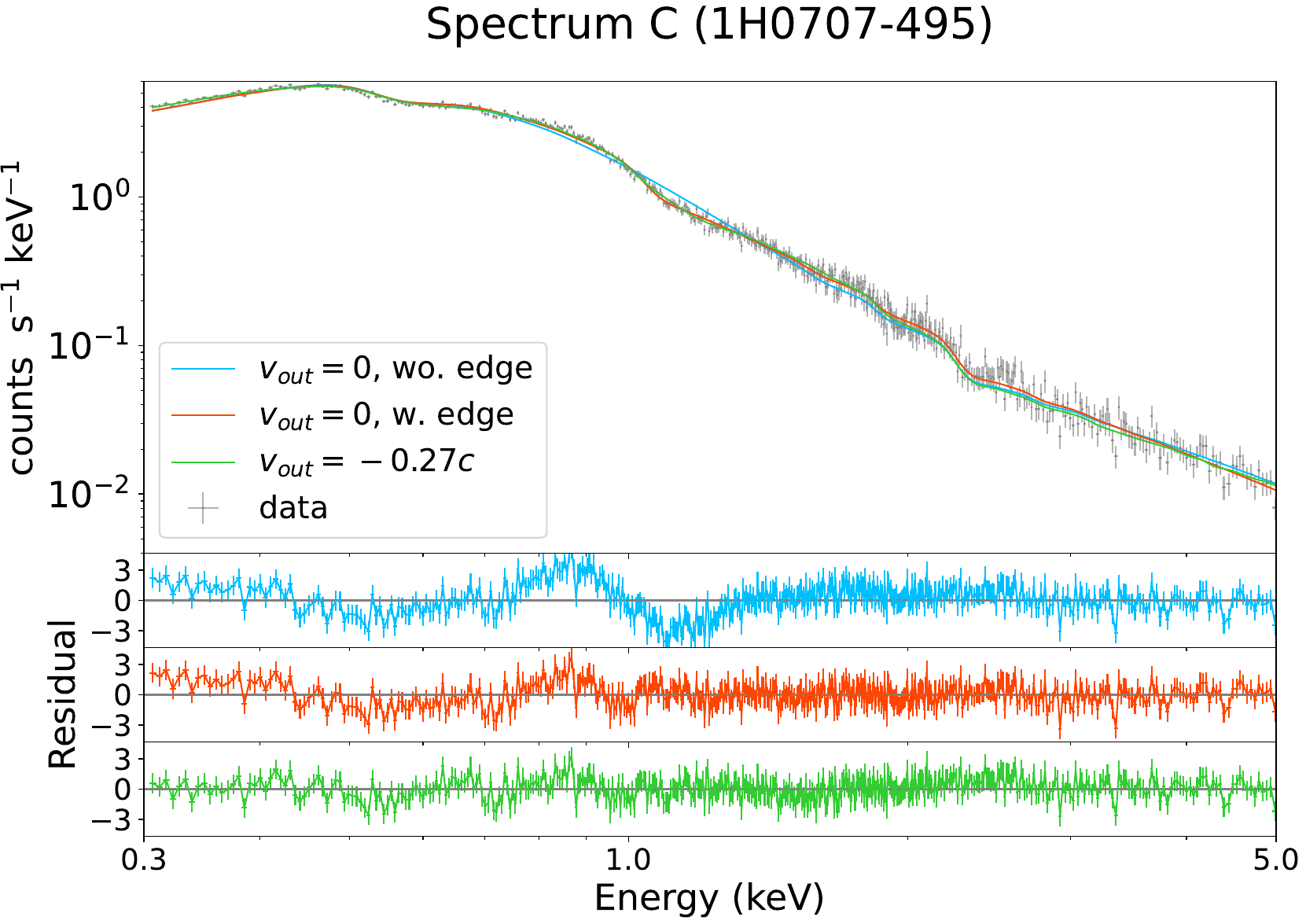}}
\vspace{3mm}
 \caption{Spectral fitting results for the spectrum C of 1H~0707$-$495 in 2008 in 0.3--5.0~keV to compare different models. The light-blue and red lines show the best-fit models without/with artificial edges around 1.0~keV when the clump velocity is fixed at zero, while the green line shows the best-fit model when the clump velocities are blueshifted at 0.27 $c$.\\
 {Alt text: The main panel shows the X-ray spctra. The lower three panels show the residual for different models.}} 
	 \label{fig:modelcomp_vel0_edge}
\end{figure}

In Figure~\ref{fig:modelcomp_vel0_edge}, we show spectral fitting results for 1H~0707$-$495 in the 0.3--5.0~keV range with three different models.
The light-blue color shows the best-fit models and residuals with the stationary clumpy absorbers, which gives large residuals around 1~keV.
The red gives the model with an artificial $\sim$1 keV edge to explain the residuals. 
The green is the best-fit model with the {\em outflowing}\/ clumpy absorbers with 0.27 $c$ without 
an inexplicable edge.
This result strongly suggests that the residual feature was caused by the energy shift of the outflowing clumpy absorbers at extremely high velocities.
Consequently, the successful reproduction of this complex spectral feature provides robust support for the physical presence of the ``less-ionized clumpy UFOs'' proposed in this study, demonstrating the effectiveness of our spectral-ratio analysis.

\section{Conclusion} \label{Chapter6} 
In this paper, we systematically analyzed \textit{XMM-Newton} data of Seyfert~1 galaxies to constrain
the partial covering clump velocities using the novel {\em spactral-ratio model fitting method}\/ \citep{Midooka22a,midooka23}. 
Among the 12 sources we studied, 1H~0707$-$495, PDS~456, and Mrk 335, exhibited significant
dip/clif structures at $\sim$1 keV in their spectral-ratios, which were explained by introducing the
mildly ionized partial absorbing clumps. The clump parameters, 
$\xi$, $N_\mathrm{H}$, and $v_\mathrm{out}$ were constrained from the spectral-ratio model fitting.
Consequently, we discovered that the clumpy absorbers have unexpectedly high outflow velocities ($> 0.1 c$) for 1H~0707$-$495 and PDS~456.
Furthermore, the clump velocities get higher with increasing X-ray fluxes in two of four data sets (Mrk 335 and 1H 0707--495 in 2010), and the remaining two datasets also agree with the positive correlation.
This correlation was also observed in IRAS~13224$-$3809.
Assuming that the  mass of the outflowing gas is constant, the flux dependence of the outflow velocity suggests that both the UFOs and clumps are radiatively driven and have the same origin, which is consistent with the ``hot inner and clumpy outer wind scenario'' proposed by previous studies \citep[e.g., ][]{Mizumoto19}.

In addition, our outflowing clumpy model presents a plausible explanation of the energy spectra.
Previous spectral studies of 1H~0707$-$495 based on conventional models required a strong unexplained absorption edge at around 1~keV. Our model explains this residual without such an inexplicable edge, demonstrating that the residual feature is caused by the blueshift of the outflowing clumpy absorbers.

Based on the comparison of the UFO velocities and the clump velocities,
we found that the clump velocity appears comparable to the fastest UFO component,
or even faster than the UFO velocities.
Given the assumption that the UV--X-ray flux variation directly contributes to the increase in the
kinetic energy of the disk wind, 
the UFOs are, on one hand, considered to be driven by the UV-dominant continuum
radiation. On the other hand, the even faster clumpy absorbers may not be explained by
the continuum-driven scenario alone.
Thus, this study, for the first time, proposes the possibility that the line-driven scenario may contribute to the additional acceleration of the clumpy absorbers.

Our analysis probed the outflowing velocity of the X-ray absorbers and their physical origins.
However, our results obtained  with the CCD instruments remain uncertain of the UFO velocities.
Future X-ray observatories such as \textit{XRISM} 
and \textit{Athena} will enable to provide extremely high-resolution spectroscopy up to a higher energy band than that of \textit{XMM-Newton}.
Such high-quality data will precisely determine the UFO velocities and unambiguously confirm our conjecture that
the clump velocities are comparable to or even faster than the UFO velocities.
If the clumps are accelerated by the line force as suggested in the present study, it is considered that more momentum and/or energy are transported to the interstellar medium of host galaxies than those estimated only from the UFO velocities.
Therefore, the present study is expected to be a stepping stone toward a better understanding of the energy transport from the AGN innermost region to the host galaxy and the co-evolution of SMBHs and host galaxies.

\section*{Supplementary data} 

The following supplementary data are available at PASJ online: E-table 1, E-figure 1, E-figure 2, and E-figure 3.

\begin{ack}
This study was based on observations obtained with \textit{XMM-Newton}, ESA science missions with instruments and contributions directly funded by ESA Member States and NASA. 
\end{ack}

\section*{Funding}
This research was supported by JSPS Research Fellow Grant Number JP20J20809 (TM), JSPS KAKENHI Grant Number JP21K13958 (MM), and Yamada Science Foundation (MM). 

\section*{Data availability} 
The \textit{XMM-Newton} data underlying this article are available in the XMM-Newton Science Archive (XSA). All data reduction tools are available as the FTOOLS package maintained by NASA/GFSC or the SAS package maintained by ESA. 


\bibliography{00_main}
\bibliographystyle{aasjournal}

\end{document}